\documentclass[structabstract]{aa}  
\usepackage{graphicx}
\usepackage{lscape}
\usepackage{amsmath}
\usepackage{txfonts}
\usepackage{url}
\usepackage{natbib}

\begin{document}
   \title{Grids of stellar models with rotation}

   \subtitle{I. Models from 0.8 to 120 $M_{\sun}$ at solar metallicity (\textit{Z} = 0.014)}

   \author{S. Ekstr\"om\inst{1},
         C. Georgy\inst{2},
         P. Eggenberger\inst{1},
         G. Meynet\inst{1},
       N. Mowlavi\inst{1},
        A. Wyttenbach\inst{1},\\
         A. Granada\inst{1,3},
        T. Decressin\inst{1},
        R. Hirschi\inst{4,5},
        U. Frischknecht\inst{4,6},
        C. Charbonnel\inst{1,7},
        \and
          A. Maeder\inst{1}}

   \authorrunning{Ekstr\"om et al.}

   \institute{Geneva Observatory, University of Geneva, Maillettes 51, CH-1290 Sauverny, Switzerland
          \and Centre de recherche astrophysique, Ecole Normale Sup\'erieure de Lyon, 46, all\'ee d'Italie, F-69384 Lyon cedex 07, France
          \and Instituto de Astrof\'isica La Plata, CCT La Plata, CONICET-UNLP, Paseo del Bosque S/N, La Plata, Buenos Aires, Argentina
          \and Astrophysics group, EPSAM, Keele University, Lennard-Jones Labs, Keele, ST5 5BG, UK
          \and Institute for the Physics and Mathematics of the Universe, University of Tokyo, 5-1-5 Kashiwanoha, Kashiwa, 277-8583, Japan
          \and Dept. of Physics, University of Basel, Klingelbergstr. 82, CH-4056, Basel, Switzerland
          \and IRAP, UMR 5277 CNRS and Universit\'e de Toulouse, 14, Av. E.Belin, 31400 Toulouse, France}

   \date{Received ; accepted }

 \abstract
   {} 
   {Many topical astrophysical research areas, such as the properties of planet host stars, the nature of the progenitors of different types of supernovae and gamma ray bursts, and the evolution of galaxies, require complete and homogeneous sets of stellar models at different metallicities in order to be studied during the whole of cosmic history. We present here a first set of models for solar metallicity, where the effects of rotation are accounted for in a homogeneous way.}
   {We computed a grid of 48 different stellar evolutionary tracks, both rotating and non-rotating, at $Z=0.014$, spanning a wide mass range from 0.8 to 120 $M_{\sun}$. For each of the stellar masses considered, electronic tables provide data for 400 stages along the evolutionary track and at each stage, a set of 43 physical data are given. These grids thus provide an extensive and detailed data basis for comparisons with the observations. The rotating models start on the zero-age main sequence (ZAMS) with a rotation rate $\upsilon_\text{ini}/\upsilon_\text{crit}=0.4$. The evolution is computed until the end of the central carbon-burning phase, the early asymptotic giant branch (AGB) phase, or the core helium-flash for, respectively, the massive, intermediate, and both low and very low mass stars. The initial abundances are those deduced by Asplund and collaborators, which best fit the observed abundances of massive stars in the solar neighbourhood. We update both the opacities and nuclear reaction rates, and introduce new prescriptions for the mass-loss rates as stars approach the Eddington and/or the critical velocity. We account for both atomic diffusion and magnetic braking in our low-mass star models.}
   {The present rotating models provide a good description of the average evolution of non-interacting stars. In particular, they reproduce the observed main-sequence width, the positions of the red giant and supergiant stars in the Hertzsprung-Russell (HR) diagram, the observed surface compositions and rotational velocities. Very interestingly, the enhancement of the mass loss during the red-supergiant stage, when the luminosity becomes supra-Eddington in some outer layers, help models above 15-20 $M_{\sun}$ to lose a significant part of their hydrogen envelope and evolve back into the blue part of the HR diagram. This result has interesting consequences for the blue to red supergiant ratio, the minimum mass for stars to become Wolf-Rayet stars, and the maximum initial mass of stars that explode as type II-P supernovae.}
   {}
 

   \maketitle
\section{Introduction}

Even if stellar physics has already a well-established history, which has allowed this area of research to reach a high level of sophistication, many surprises and challenges continue to stimulate ongoing studies. For instance, \citet{asp05} proposed a revision of the present-day surface solar abundances which restores the compatibility of the composition of our star with the one observed in B-type stars in its vicinity. However these abundances pose some difficulties when attempting to explain the helioseismic observations, which may indicate that some pieces of physics are still missing in solar models \citep[e.g.][and references therein]{turck11}. New mass-loss rates (in general lower than the previous ones) have been proposed for the main-sequence (MS)  and Wolf-Rayet (WR) phases \citep{bouret05,fullerton06} making it more difficult to explain for instance the formation of WR stars, at least in the framework of the single star scenario. Changes in the abundances are observed at the surface of stars, which implies that mixing processes operate in radiative zones \citep[see e.g.][]{przybilla2010}. The origin of the variation with metallicity in the number ratio of blue to red supergiants still lacks a general explanation \citep{meyl82,langmaed95,eggenmm02}. These have been only a few illustrations the fact that, although many aspects of the evolution of stars are well understood, important problems remain unsolved. 

Significant progress is now achievable using observations. Multi-object spectrographs can acquire data for large numbers of stars, unveiling an unprecedentedly detailed view of how some stellar features vary as a function of mass, age, and metallicity.

Techniques such as asteroseismology, spectropolarimetry or interferometry provide complementary insight, making it possible to obtain information on features such as the size of the convective cores, the internal rotation laws, the surface magnetic fields, and the shapes of the stars and their winds, which were completely out of reach only a few years ago \citep{cunha07,walder2011}.

In the near future, the ESA new satellite GAIA will provide data for large fractions of stars in our Galaxy to help enhance our knowledge of its evolutionary history \citep[see e.g. ][]{Cacciari2009}.

Stellar models play a crucial role in extracting from observations the physical quantities needed to enhance our knowledge. Therefore, improvements of the observing facilities should be accompanied by improvements in the stellar models. In the past decade, the effects of axial rotation have been studied extensively by many groups \citep[][Chieffi \& Limongi, in prep.]{HL00,mm5,palacios03,TC05,DP07}. It does appear that rotation has an impact on all outputs of stellar models \citep{mm5,mm2010}, such as lifetimes, evolutionary scenarios, or nucleosynthesis, thus on many outputs of both population synthesis models and chemical evolution models of star clusters \citep{decr07}, starbursts, and galaxies \citep{chiap06,vazquez07}.

Grids including rotation have so far been focused on some limited mass intervals. Since the grids published by the different research groups differ in terms of their input physical  ingredients (overshooting, mass loss, nuclear reaction rates,...) or the way in which the effects of rotation are taken into account (expressions of the diffusion coefficients, treatment of the angular momentum equation), their results cannot be combined to provide a coherent and homogeneous picture. In the present series of papers, our ambition is to construct a new database of evolutionary models incorporating the main improvements made in recent years to stellar ingredients and accounting in particular for the effects of axial rotation. We take care to compute the models in a homogeneous way allowing us to provide a coherent and consistent view over an extended mass and metallicity range. In this first paper, we focus on solar metallicity models.
 
The paper is structured as follows: the physical ingredients of the models are discussed in Sect.~\ref{SecPhymod}. The computed stellar models and the electronic tables, which we make publicly available via a web page, are presented in  Sect.~\ref{SecTables}. The properties of the non-rotating (or initially slowly rotating) tracks are discussed in Sect.~\ref{SecResultsNOROT}, while the rotating tracks are presented in Sect.~\ref{SecResultsROT}. We briefly conclude in Sect.~\ref{SecDiscu}.

\section{Physical ingredients of the models \label{SecPhymod}}
\subsection{Abundances, opacities, and equation of state \label{SubSecAbund}}

The initial abundances of H, He, and metals are set to $X=0.720$, $Y=0.266$, and $Z=0.014$. We obtained these initial abundances by calibrating a 1 $M_{\sun}$ model including atomic diffusion to reproduce the abundances measured at the surface of the Sun, as well as the solar radius and luminosity after 4.57\,Gyr. The mixture of heavy elements is assumed to be that of \citet{asp05} except for the Ne abundance, which is taken from \citet{cunha06}. Interestingly, the abundances correspond to the measured abundances in young massive stars of the solar neighbourhood \citep[see][their Table 4]{asplund09ARAA}. The detailed element abundances are presented in Table~\ref{TabAbundini}. We note that the isotopic ratios are taken from \citet{Lodders2003}.
\begin{table}
\caption{Chemical initial abundances of the models given in $\log (X/H)+12$ units and in mass fraction.}
\label{TabAbundini}
\centering
\begin{tabular}{r r| r r}
\hline\hline
Element & Isotope & $\log (X/H)+12$ & mass fraction \\
\hline
    H   &  & 12.00 & 7.200e-01 \\
          &  &             &                    \\
    He &  & 10.93 & 2.660e-01 \\
          & \element[][3]{He} & & 4.415e-05 \\
          & \element[][4]{He} & & 2.660e-01 \\
          &  &             &                    \\
    C   &  &   8.39 & 2.311e-03 \\
          & \element[][12]{C} & & 2.283e-03 \\
          & \element[][13]{C} & & 2.771e-05 \\
          &  &             &                    \\
    N   &  &   7.78 & 6.614e-04 \\
          & \element[][14]{N} & & 6.588e-04 \\
          & \element[][15]{N} & & 2.595e-06 \\
          &  &             &                    \\
    O   &  &   8.66 & 5.734e-03 \\
          & \element[][16]{O} & & 5.718e-03\\
          & \element[][17]{O} & & 2.266e-06\\
          & \element[][18]{O} & & 1.290e-05\\
          &  &             &                    \\
    Ne &  &  8.11 & 2.029e-03 \\
          & \element[][20]{Ne} & & 1.877e-03\\
          & \element[][22]{Ne} & & 1.518e-04\\
          &  &             &                    \\
    Na & &  6.17 & 2.666e-05 \\
          & \element[][23]{Na} & & 2.666e-05 \\
          &  &             &                    \\
    Mg & &  7.53 & 6.459e-04 \\
          & \element[][24]{Mg} & & 5.035e-04\\
          & \element[][25]{Mg} & & 6.641e-05\\
          & \element[][26]{Mg} & & 7.599e-05\\
\hline
\end{tabular}
\end{table}

The opacities are generated with the OPAL tool\footnote{\url{http://adg.llnl.gov/Research/OPAL/opal.html}} \citep[based on][]{opa96} for this particular mixture. They are complemented at low temperatures by the opacities from \citet{ferg05} adapted for the high Ne abundance.

Solar-type models with $M < 1.25\,M_{\sun}$ are computed with the OPAL equation of state \citep{rog02}. For the higher mass models, the EOS is that of a mixture of perfect gas and radiation, and account for partial ionisation in the outermost layers, as in \citet{schaller92}, and for the partial degeneracy in the interior in the advanced stages.

\subsection{Nuclear reaction rates \label{SubSecReac}}

The nuclear reaction rates are generated with the NetGen tool\footnote{\url{http://www-astro.ulb.ac.be/Netgen/}}. They are taken mainly from the Nacre database \citep{ang99}, although some have been redetermined more recently and updated. We list them below, with a comparison to NACRE values and a short description of the effects on stellar evolution.

The rate of \element[][14]{N}($p,\gamma$)\element[][15]{O} is taken from \citet{mukh03}. It is about half the NACRE value for temperatures below $10^{8}$ K, and compares well with other determinations such as \citet{AngDesc01} or \citet{luna06}. This reaction is the slowest of the CNO cycle. In the low-mass domain, the effects of lowering this rate have been studied by \citet{imbr04} and \citet{weiss05}. They describe a slower H-burning process, and shallower temperature profiles leading to a more extended and slightly hotter core. The turn-off point is shifted towards a higher luminosity. In the intermediate-mass domain, the studies of \citet{herw06} and \citet{weiss05} show that with lower rates, the MS evolution occurs at a higher luminosity, and that later, the blue loops during core He burning get significantly shorter.

The rate of \element[][4]{He}($\alpha \alpha,\gamma$)\element[][12]{C} is taken from \citet{fyn05}. This rate is around twice the NACRE value below $25\cdot10^{6}$ K, slightly lower than NACRE between $25\cdot10^{6}$ and $10^{8}$ K, similar to NACRE between $10^{8}$ and $10^{9}$ K, and a tenth of the NACRE value above $10^{9}$ K. The effects of this new rate on stellar evolution were studied by \citet{weiss05} and \citet{herw06} for intermediate-mass stars. The differences are extremely small in this mass range. In the massive star domain, a lower rate in the temperature range of core He burning is expected to lead to a slightly larger core at the end of the evolution, with a low C/O ratio, and slightly lower \element[][12]{C} yields \citep{finestruc2010}. The uncertainties in this rate, as well as the \element[][12]{C}($\alpha,\gamma$)\element[][16]{O} rate, is shown by \citet{tur09,tur10} to strongly affect the production of the medium-weight nuclei and the the weak $s$-process nuclei.

The rate of \element[][12]{C}($\alpha,\gamma$)\element[][16]{O} is taken from \citet{kunz02}. It is around 0.6-0.8 times the NACRE value below $6\cdot10^{8}$ K, and around 1.1-1.4 times the NACRE value above this temperature. Both \citet{ww93} and \citet{imbr01} explored the effects of varying this rate. A higher rate in the He-burning temperature range leads to larger cores, lower \element[][12]{C}, and higher \element[][16]{O} yields.

The rate of \element[][22]{Ne}($\alpha,n$)\element[][25]{Mg} is taken from \citet{jaeg01}. It is around 0.6 times the NACRE value below $6\cdot10^{8}$ K. While there is little effect on stellar evolution, the ratio of the isotopes of Mg are modified \citep{karak06}, and less \element[][25]{Mg} and \element[][26]{Mg} is produced. This reaction is a neutron source for the $s$-process, and \citet{the07} studied the consequences of reducing its rate, which they found would lead to a significant reduction in the $s$-process efficiency during core He burning.

We note that the expected impact of the updated nuclear reaction rates is small and dwarfed by the effect of other modifications.

The models of massive stars ($M > 9 M_{\sun}$) were computed by incorporating the NeNa-MgAl cycle. For this sub-network, the NACRE rates were also used, except for two reactions that were updated: \element[][21]{Ne}($p,\gamma$)\element[][22]{Na} \citep{iliad01} and \element[][22]{Ne}($p,\gamma$)\element[][23]{Na} \citep{hale02}. The updated reaction rates do not have a significant effect on the structure and evolution of the models. The effects of rotation and these new rates on the nucleosynthesis, in particular that of the $s$-process, will be discussed in a forthcoming paper (Frischknecht et al, in prep.).

The energy loss in plasma, pair, and photo-neutrinos processes are taken from \citet{itoh89} and \citet{itoh96}.

\subsection{Convection and diffusion \label{SubSecConv}}

\begin{figure}
\includegraphics[width=.48\textwidth]{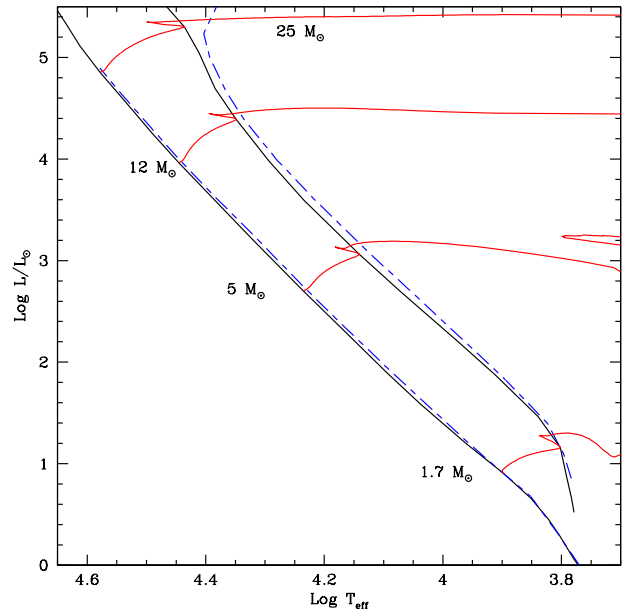}
   \caption{Comparison of the MS band between the present rotating models (solid black line) and the non-rotating models of \citet{schaller92} (long-short dashed blue line).}
    \label{MS}
\end{figure}

The convective zones are determined with the Schwarzschild criterion. For the H- and He-burning phases, the convective core is extended with an overshoot parameter $d_\text{over}/H_P=0.10$ from 1.7 $M_{\sun}$ and above, 0.05 between 1.25 and 1.5 $M_{\sun}$, and 0 below. If $d_\text{over}$ exceeds the dimension of the convective core $R_\text{cc}$, then the total extension of the convective core is given by $R_\text{cc}(1+d_\text{over}/H_P)$. This procedure avoids any core-extension amplitude larger than the radius of the initial core. The value of the overshoot parameter was calibrated in the mass domain $1.35-9\, M_{\sun}$ to ensure that the rotating models closely reproduce the observed width of the MS band.

In Fig.~\ref{MS}, we compare the MS bandwidth of the present models with that in the models published in \citet{schaller92}. Our new models concur to a slightly narrower MS width than \citet{schaller92}, which provided a good fit to the observed MS width \citep[see the discussion in][]{schaller92}. However, since stars have a distribution of initial rotation velocities,  stars at the end of the MS phase will scatter around the limit shown by our moderately rotating models, more rapid rotators lying beyond it and slightly enlarging the MS width. A more detailed discussion and a comparison of the MS width obtained from our models with observations are provided in Sect.~\ref{SubSecMSWidth}. We note for now that the combined effects of rotational mixing and of an overshoot of 0.1 mimic the effects obtained in models with no rotational mixing and a stronger overshoot of 0.2 (used in the 92's models). This is of course expected because rotational mixing also contributes to making the convective cores larger at a given evolutionary stage. This effect was discussed in \citet{talon97}.

\begin{figure}
\includegraphics[width=.48\textwidth]{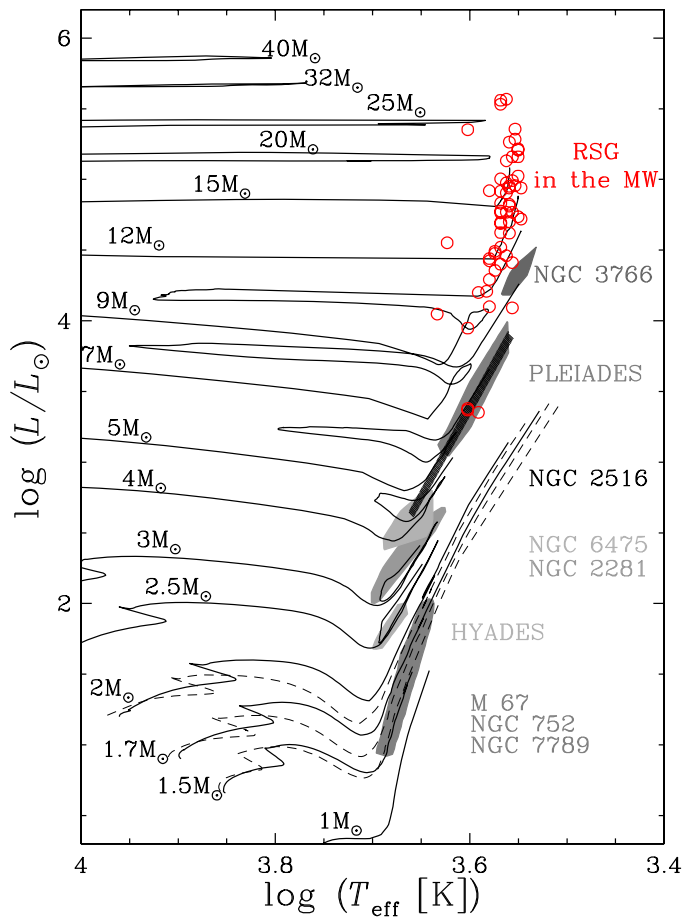}
   \caption{Evolutionary tracks for rotating models in the red part of the HR diagram. In the low mass range, a few non-rotating tracks (dashed lines) are shown. The grey shaded areas indicate the observations in clusters and associations as given in \citet{MM89} as well as the position of Galactic red supergiants obtained by \citet[red circles]{levesque05}.}
    \label{rsg}
\end{figure}

The outer convective zone is treated according to the mixing length theory, with a solar calibrated value for the mixing-length parameter of the low-mass stars ($\alpha_\text{MLT} \equiv \ell/H_P=1.6467$). For more massive stars, {\it i.e.} for stars with $M>1.25\, M_{\sun}$, the difference in the EOS implies a slightly lower value for this parameter: $\alpha_\text{MLT}=1.6$. In Fig.~\ref{rsg}, we see that the present models closely reproduce the positions of both the red giant branch and the red supergiants, supporting the choices made for the value of the mixing length.

For the most luminous models, the turbulence pressure and acoustic flux need to be included in the treatment of the envelope. As in \citet{schaller92}, this is done according to \citet[see also \citealt{MaederBook09}, Sect.~5.5]{Maeder87}, using a mixing length taken on the density scale: $\alpha_\text{MLT}=\ell/H_\rho = \ell (\alpha - \delta \nabla) / H_P = 1$. The use of $\ell/H_\rho$ instead of $\ell/H_P$ removes the risk of having an unphysical density inversion in the envelope \citep{StoChi73}. The side effect of this treatment is that the redward extension of the tracks in the Hertzsprung-Russell (HR) diagram is reduced by 0.1-0.2 dex in $T_\text{eff}$ \citep[see][]{MM87}, which is why we apply it only to the models with $M\geq40\,M_{\sun}$, which do not extend to the extreme red part of the HR diagram.

Low-mass stellar models with $M < 1.25\,M_{\sun}$ are computed with the effect of atomic diffusion caused by both concentration and thermal gradients \citep[see][ for more details]{egg08}.

\subsection{Rotation \label{SubSecRot}}

The treatment of rotation was developed in a series of papers published previously by our group \citep{Maed99,Maed97,mm6,mz98,mm1,mm5}. The interested reader may refer to these papers to get the full developments. We recall here the main aspects of this treatment. 

\subsubsection{Diffusion coefficients, meridional circulation}

We first apply the shellular-rotation hypothesis, which postulates that in differentially rotating stars, the horizontal turbulence, \textsl{i.e.} the turbulence along an isobar, is very strong. This is expected because there is no restoring force in that direction, as the buoyancy force (the restoring force of the density gradient) acts in the vertical direction. \citet{Za92} relates the diffusion coefficient to the viscosity caused by horizontal turbulence
\begin{equation}
D_\text{h} \approx \nu_\text{h} = \frac{1}{c_\text{h}}\ r\ \left| 2\,V(r) - \alpha\,U(r) \right|,
\label{EqDh}
\end{equation}
where $c_\text{h}$ is a constant of the order of 1 (here taken $=1$), $V(r)$ is the horizontal component of the meridional circulation velocity, $U(r)$ its vertical component (see below), and in this expression $\alpha = \frac{1}{2} \frac{\text{d} \ln (r^2 \bar{\Omega})}{\text{d} \ln r}$.

According to the von Zeipel theorem \citep{vZ24}, a rotating star cannot be locally in both hydrostatic and radiative equilibrium at the same time. This drives a large-scale circulation, known as the \textit{meridional circulation} \citep[or \textit{Eddington-Sweet circulation}, ][]{Edd25,Sweet50}. Its vertical component is to first order $u(r,\theta)=U(r)\,P_2(\cos \theta)$, where $P_2(x)$ is the second Legendre polynomial\footnote{The Legendre polynomials are a set of orthogonal functions used to solve Laplace's equation $\nabla^2\Phi=0$ in spherical coordinates when there is azimuthal symmetry $\partial\varphi=0$. The second-order Legendre polynomial is $P_2(x)=\frac{1}{2}\left( 3x^2-1\right)$.}. The formulation of the radial amplitude $U(r)$ was determined by \citet{Za92} and \citet{mz98} to be
\begin{align}
U(r)= & \frac{P}{\rho g C_P T}\frac{1}{\left[ \nabla_\text{ad} - \nabla_\text{rad} + (\varphi/\delta) \nabla_\mu \right]} \nonumber \\
 & \cdot \left( \frac{L}{M_\star} \left[ E_\Omega^\star + E_\mu \right] + \frac{C_P}{\delta} \frac{\partial \Theta}{\partial t} \right),
\label{EqUr}
\end{align}
where $C_P$ is the specific heat at constant pressure, $\nabla_\text{ad}=\frac{P \delta}{\rho T C_P}$ is the adiabatic gradient, $M_\star=M \left(1 - \frac{\Omega^2}{2\pi g \rho_\text{m}} \right)$, and $\Theta=\tilde{\rho}/\bar{\rho}$ is the ratio of the variation of the density to the average density on an equipotential. Both $\varphi$ and $\delta$ arise from the generic equation of state $\frac{\text{d} \rho}{\rho} = \alpha \frac{\text{d} P}{P} + \varphi \frac{\text{d} \mu}{\mu} - \delta \frac{\text{d} T}{T}$, and $E_\Omega^\star$ and $E_\mu$ are terms that depend on the $\Omega$- and $\mu$-distributions respectively \citep[see][for details on these expressions]{mz98}.

We can then define an effective diffusion coefficient, $D_\text{eff}$, which combines the effects of the horizontal diffusion and that of the meridional circulation as \citep{chabz92}
\begin{equation}
D_\text{eff} = \frac{1}{30} \frac{\left| r\ U(r) \right|^2}{D_\text{h}}.
\label{EqDeff}
\end{equation}

Differential rotation induces shear turbulence at the interface of layers that have different rotational velocities. A layer remains stable if the excess kinetic energy due to the differential rotation does not overcome the buoyancy force (known as the \textit{Richardson criterion}). Shear arises when the thermal dissipation reduces the buoyancy force. The coefficient of diffusion by shear turbulence is determined by \citet{Maed97} to be
\begin{align}
D_\text{shear} =& \frac{K}{\frac{\varphi}{\delta}\nabla_\mu + \left( \nabla_\text{ad} - \nabla_\text{rad} \right)} \nonumber \\
       & \cdot \frac{H_P}{g\delta} \left[f_\text{energ} \left(\frac{9\pi}{32}\ \Omega\ \frac{\text{d} \ln \Omega}{\text{d} \ln r} \right)^2 - \left(\nabla' - \nabla \right) \right],
\label{EqShear}
\end{align}
where $K=\frac{4acT^3}{3\kappa \rho^2 C_P}$ is the thermal diffusivity, $f_\text{energ}$ is the fraction of the excess energy in the shear that contributes to mixing (here taken $=1$), and $(\nabla' - \nabla)$ is the difference between the internal non-adiabatic gradient and the local gradient. We note that $(\nabla' - \nabla)$ can be neglected in most cases.

\subsubsection{Transport mechanisms}

The transport of angular momentum inside a star is implemented following the prescription of \citet{Za92}. This prescription was complemented by \citet{TZ97} and \citet{mz98}. In the radial direction, it obeys the equation
\begin{equation}
\rho \frac{\text{d}}{\text{d}t} \left( r^2 \bar{\Omega} \right)_{M_r} =
   \frac{1}{5r^2} \frac{\partial}{\partial r} \left( \rho r^4 \bar{\Omega} U(r) \right)
   + \frac{1}{r^2} \frac{\partial}{\partial r} \left( \rho D r^4 \frac{\partial \bar{\Omega}}{\partial r} \right).
\label{EqTranspAng}
\end{equation}
The first term on the right hand side of this equation is the divergence of the \textit{advected} flux of angular momentum, while the second term is the divergence of the \textit{diffused} flux. The coefficient $D$ is the total diffusion coefficient in the vertical direction, taking into account the various instabilities that transport angular momentum (convection, shears).

\citet{chabz92} show that the horizontal turbulence competes efficiently with the advective term of the meridional circulation for transporting the chemical species. The horizontal flow tends to homogenise the layer in such a way that the transport of chemical species by both meridional circulation and horizontal turbulence can be computed as a diffusive process with the coefficient $D_\text{eff}$ calculated in Eq.~\ref{EqDeff}. The change in the abundance for a given chemical element $i$ in the shell with Lagrangian coordinate $r$ is thus:
\begin{equation}
\rho \frac{\text{d}X_i}{\text{d}t} = \frac{1}{r^2} \frac{\partial}{\partial r} \left( \rho r^2 \left[ D + D_\text{eff} \right] \frac{\partial X_i}{\partial r} \right) + \left( \frac{\text{d}X_i}{\text{d}t}\right)_\text{nucl},
\label{EqChem}
\end{equation}
where $D$ is the same as in Eq.~\ref{EqTranspAng}. The last term accounts for the change in abundance produced by nuclear reactions.

\subsubsection{Angular momentum conservation \label{SubSecAngCons}}

The final angular momentum content of the star (particularly the core content) is a key quantity for determining its final state \citep[supernova, hypernova, gamma-ray burst, see e.g.][]{Nomoto2005,YLN06}. The conservation of  angular momentum during the whole stellar evolution is therefore of prime importance. In the Geneva code, the angular momentum conservation is checked throughout the evolution of the star. During the computation of the models, there are two sources of variations in the total amount of angular momentum:
\begin{itemize}
\item A numerical variation, owing to unavoidable inaccuracies in the resolution of the advection-diffusion equation of the angular momentum transport and the structure of the code itself \citep{Kippenhahn1967a}: the envelope\footnote{In our models, the envelope is the region above a given mass coordinate in which the luminosity is considered as constant, and where partial ionisation is accounted for.} is "floating" over the interior layers, and is assumed to rotate at the same angular velocity than the most superficial shell of the interior. Its angular momentum content is therefore imposed by what happens in the interior. For the total angular momentum of the star to remain constant, a correction needs to be applied (see below).
\item An angular momentum loss due to the mass loss of the rotating star, where the stellar winds carry away some amount of momentum.
\end{itemize}

Knowing the mass-loss rate $\dot{M}$ of the star (as a function of stellar parameters such as the luminosity, the effective temperature, and the chemical composition of the surface, cf. Sect.~\ref{SubSecMdotRad}), it is possible to compute the amount of angular momentum $\Delta\mathcal{L}_\text{winds}$ carried away by the stellar winds during the current time step $\Delta t$
\begin{equation}
\Delta\mathcal{L}_\text{winds} = \frac{2}{3}\dot{M}\Omega_\text{S}r_*^2\Delta t,
\label{EqDeltaLWinds}
\end{equation}
where $\Omega_\text{S}$ is the angular velocity of the surface, and $r_*$ the stellar radius. We also assumed that the mass loss is spherically symmetric, even if some anisotropy may develop. \citet{georgy11} demonstrated that this assumption leads to only small errors, particularly for stars that remain far from the critical rotation\footnote{The critical velocity is reached when the gravitational acceleration is exactly counterbalanced by the centrifugal force. In the framework of the Roche model, one has $\upsilon_\text{crit}=\sqrt{\frac{2}{3}\frac{GM}{R_\text{pb}}}$, where $R_\text{pb}$ is the polar radius at the critical limit.}, as in this work. Starting from the total angular momentum $\mathcal{L}_\text{ini}$ of the model at the time step $n-1$, the angular momentum conservation ensures that the expected angular momentum at the end of the $n^\text{th}$ time step should be
\begin{equation}
\mathcal{L}_\text{fin}^\text{exp} = \mathcal{L}_\text{ini} - \Delta\mathcal{L}_\text{winds}.
\end{equation}
As discussed above, the final angular momentum $\mathcal{L}_\text{fin}^\text{ob}$ obtained after the whole computation of the structure differs usually from the expected one $\mathcal{L}_\text{fin}^\text{exp}$. We thus need to correct the angular momentum of the model by an amount $\Delta\mathcal{L}_\text{corr} = \mathcal{L}_\text{fin}^\text{ob} - \mathcal{L}_\text{fin}^\text{exp}$. The correction for the angular velocity of the $i^\text{th}$ shell is of the form of $\Omega_i^\text{corr} = \Omega_i^\text{ob}\left(1+q_\text{corr}\right)$ where $\Omega_i^\text{ob}$ is the angular velocity of the $i^\text{th}$ shell obtained by the numerical computation, and $q_\text{corr}$ the correction factor, given by
\begin{equation}
q_\text{corr} = -\frac{\Delta\mathcal{L}_\text{corr}}{\mathcal{L}_\text{e}+\sum_{i=1}^{N_\text{corr}}\mathcal{L}_i},
\end{equation}
where $\mathcal{L}_\text{e}$ and  $\mathcal{L}_i$ are the angular momenta of the envelope and  the $i^\text{th}$ shell, respectively. A detailed description of the method can be found in \citet{Georgy10phdt}.

The choice of $N_\text{corr}$ is fixed, and corresponds roughly to the zone that can be reached by the advection-diffusion of the angular momentum during one characteristic time step.

\subsection{Magnetic field}

We accounted for neither a dynamo mechanism in the stellar interior, nor any strong fossil field that would impose solid body rotation. Our present knowledge of the dynamo theory is too uncertain \citep{zbm07}, and the observational constraints from spectropolarimetry remain too weak to account for these effects in a reliable way. According to \citet{mmB3}, an internal magnetic field produces a strong internal coupling, which keeps the star at a higher surface rotational velocity throughout the whole MS. This may provide an interesting test of whether a significant internal magnetic coupling is present.

We note that the main reasons for introducing the effects of internal magnetic fields that would enforce solid body rotation during the MS phase come from essentially two directions: to provide first an explanation for the flat rotational profile in the Sun \citep{brown89,koso97,couvidat03,rotmag4}, and second, a more efficient mechanism for slowing down the cores to explain the observed rotation rates of young pulsars \citep{hws05} and white dwarfs \citep[WD,][]{suijs08}. The problems of the rotational profile of the Sun and the spin rate of pulsars might not be linked, since the physics of solar-type stars have peculiarities that are not shared by the massive ones. We note that in both questions (solar profile and pulsar rotation), alternative models have been suggested to reconcile the models with the observations: the flat solar rotational profile could be due to angular momentum transport by gravity waves \citep{Schatzman93,zahn97,talon02,TC05}, the young pulsars may have slower rotational velocities than predicted by stellar evolution models because of some efficient braking mechanism occurring at the time of the supernova explosion \citep[for instance, the propeller mechanism, see][]{IllSun75,alpar01} or during the evolution of the neutron stars in the period preceding the time when the pulsar is observed \citep{thompson04,WH04iau215}. While kicks are not likely to occur in WD formation, surface magnetic braking (see next paragraph) might provide a solution to the problem of the low-mass WD spin rates \citep[see the 1 $M_{\sun}$ model of ][]{suijs08}. For all these reasons, we see no real compelling reasons to account for either a dynamo mechanism in the internal radiative zones or the effects of a strong fossil internal magnetic field. In support of this view are the few but steadily increasing pieces of observational evidence now coming from asteroseismology for differential rotation in B-type stars \citep{suarez07,Thoul09,Kawaler2009}.

Surface magnetic fields are however expected to produce a magnetic braking of stars that have a significant outer convective zone on the main sequence, hence magnetic braking has thus far been applied to models with $M_\text{ini} < 1.7\,M_{\sun}$. We adopt the braking law of \cite{kri97} where we calibrated the braking constant to ensure that the 1\,$M_{\odot}$ rotating model reproduces the solar surface rotational velocity after $4.57$\,Gyr. We did not account for surface magnetic braking in more massive stars, although there is, in at least one case, evidence of magnetic braking \citep{townsend2010}. Again, this effect needs to be studied in greater detail before being incorporated into the regular models and we defer a more detailed study of the consequences of this effect to a future paper.

\subsection{Mass loss \label{SubSecMdot}}

Mass loss is a key ingredient governing the evolution of stars. Including it in stellar models relies on prescriptions proposed by both observers and theorists and based on fundamental stellar parameters. The prescriptions implemented in our models are described below.

\subsubsection{Radiative mass loss \label{SubSecMdotRad}}

On the MS, stars with masses below 7 $M_{\sun}$ are computed at constant mass. Above 7 $M_{\sun}$, the radiative mass-loss rate adopted is from \citet{vink01}. In the domains not covered by this prescription, we use \citet{dJ88}.

For red (super)giants, we use the \citet{Reimers75,Reimers77} formula (with $\eta=0.5$) for stars up to 12 $M_{\sun}$. The \citet{dJ88} prescription is applied for masses of 15 $M_{\sun}$ and above, to models with $\log (T_\text{eff}) > 3.7$. For $\log (T_\text{eff}) \leq 3.7$, we perform a linear fit to the data from \citet{sylvester98} and \citet{vloon99} \citep[see][]{Crowther00}. The WR stars are computed with \citet{nuglam00} prescription, or the \citet{grafham08} recipe in the small validity domain of this prescription. In some cases, the WR mass-loss rate of \citet{grafham08} is lower than the rate of \citet{vink01}. In these cases, we use the \citet{vink01} prescription instead of that of \citet{grafham08}. Both the \citet{nuglam00} and \citet{grafham08} mass-loss rates account for some clumping effects \citep{muijres11} and are a factor of two to three lower than the rates used in the normal case of \citet{schaller92} grid.

For rotating models, we applied to the radiative mass-loss rate the correction factor described in \citet{mm6}
\begin{eqnarray}
\dot{M}(\Omega) &=& F_{\Omega}\cdot \dot{M}(\Omega=0)= F_{\Omega}\cdot \dot{M}_\text{rad}  \nonumber \\
& &\text{with} \hspace{.3cm}F_{\Omega}=\frac{(1-\Gamma)^{\frac{1}{\alpha}-1}}{\left[ 1-\frac{\Omega^2}{2\pi G \rho_\text{m}} - \Gamma \right]^{\frac{1}{\alpha}-1}},
\label{EqMdotRot}
\end{eqnarray}
where $\Gamma=L/L_\text{Edd}=\kappa L / (4\pi cGM)$ is the Eddington factor (with $\kappa$ is the electron-scattering opacity), and $\alpha$ the force multiplier parameter depending on $T_\text{eff}$.
\subsubsection{Supra-Eddington mass loss \label{SubSecMdotEdd}}

For some stellar models, particularly for massive stars ($> 15\,M_{\sun}$) in the red supergiant phase, some of the most external layers of the stellar envelope might exceed the Eddington luminosity of the star $L_\text{Edd} = 4\pi cGM/\kappa$. This is due to the opacity peak produced by the variation in the ionisation level of hydrogen beneath the surface of the star. The high opacity decreases the Eddington luminosity in these layers, possibly to fainter levels than the actual stellar luminosity, a situation that may have many consequences \citep{MaederBook09}. To account for this unstable situation (which is not solvable with our hydrostatic approach), we artificially increase the mass-loss rate of the star (computed according to the prescription described in Sect.~\ref{SubSecMdotRad}) by a factor of 3, whenever the luminosity of any of the layers of the envelope is higher than five times the Eddington luminosity.

Very interestingly, the time-averaged mass-loss rate of the 20 $M_{\sun}$ model in the red supergiant phase ({\it i.e.} when $\log(T_{\rm eff})$ is inferior to 3.65) is between $\log(\dot{M}) = -4.8$ and $-4.6\ [M_{\sun}\,\text{yr}^{-1}]$ (similar values are obtained for the rotating and the non-rotating model). These values are about one order of magnitude higher than the mass-loss rates used for the same phase in \citet{schaller92} taken from \citet{dJ88} and are compatible with the more recent mass-loss rates for red supergiants obtained by \citet{vloon05}. These new higher mass-loss rates during the red supergiant stage have important consequences for the blue to red supergiant ratio, the minimum mass of stars evolving into the Wolf-Rayet phase, and the type of supernova that these stars will produce. We briefly discuss these points in Sect.~\ref{SubSecMassiveEvol}.

\subsubsection{Mechanical mass loss}

We call mechanical mass loss, $\dot M_{\rm mech}$, the mass per unit time lost equatorially when the surface velocity of the star reaches the critical velocity, where $\dot M_{\rm mech}$ is determined by the angular momentum that needs to be lost to ensure that the surface velocity remains subcritical. The matter might be launched into an equatorial Keplerian disk, a subject that will be studied in a forthcoming paper (Georgy et al. in prep.).

The computational method used to estimate the equatorial, mechanical mass loss is detailed in \citet{Georgy10phdt}. Since the initial velocity we have chosen for this grid is moderate, none of our models reach the critical limit during the MS. Only the 9 $M_{\sun}$ model reaches it briefly during its Cepheid blue loop, losing at most $10^{-3}\,M_{\sun}$ in the process. In Georgy et al. (in prep), we will consider the case of more rapid rotators, which reach the critical limit on the MS.

\section{The stellar models and electronic tables \label{SecTables}}

Models of 0.8, 0.9, 1, 1.1, 1.25, 1.35, 1.5, 1.7, 2, 2.5, 3, 4, 5, 7, 9, 12, 15, 20, 25, 32, 40, 60, 85, and 120 $M_{\sun}$ are presented. For each mass, we computed both a rotating and a non-rotating model.

The rotating models start on the zero-age main-sequence (ZAMS) with a value of $\upsilon_\text{ini}/\upsilon_\text{crit}=0.4$. We decided to compute the rotating models at a given $\upsilon_\text{ini}/\upsilon_\text{crit}$ ratio rather than a given initial equatorial velocity because the effects of rotation are linked to this ratio, and the critical velocity $\upsilon_\text{crit}$ depends on the mass of the model. For such a widespread mass domain, a given velocity would thus correspond to very different $\upsilon/\upsilon_\text{crit}$ ratios between the different masses considered. It would even correspond to an over-critical velocity for some low-mass models. The choice of $\upsilon_\text{ini}/\upsilon_\text{crit}=0.40$ is based on the peak of the velocity distribution of young B stars in \citet[see their Fig.~6]{huang10}. This initial rotation rate corresponds to mean MS velocities of between 110 and 220 km s$^{-1}$ for the stars that are not magnetically braked ($M_\text{ini} \ge 1.7\,M_{\sun}$). These values are well within the range of those observed in the Galaxy \citep{duft06,HG06a}. 

The models are evolved up to the end of the core carbon burning ($M_\text{ini} \ge 12\,M_{\sun}$), the early asymptotic giant branch ($2.5\,M_{\sun} \le M_\text{ini} \le 9\,M_{\sun}$), or the helium flash ($M_\text{ini} \le 2\,M_{\sun}$).

Electronic tables of the evolutionary sequences are available on the web\footnote{See the webpage \url{http://obswww.unige.ch/Recherche/evol/-Database-} or the CDS database at \url{http://vizier.u-strasbg.fr/viz-bin/VizieR-2}.}. For each model, the evolutionary track is described by 400 selected data points, each point corresponding to the state of the considered star at a given age. The points of different evolutionary tracks with the same line number correspond to similar stages to facilitate the interpolation between the tracks. The points are numbered as follows:

\begin{tabbing}
\hspace*{.2cm}\=351-369:\ \ \= \kill
 \>\textbf{1:}\> \textbf{ZAMS}\\
 \>2-84:\> H burning (first part)\\
 \>85:\> minimum of $T_\text{eff}$ on the MS\\
 \>86-109:\> overall contraction phase before the end of the MS\\
 \>\textbf{110:}\> \textbf{Turn-off}\\
 \>111-189:\> HR diagram crossing and/or pre-He-b core contraction\\
 \>\textbf{190:}\> \textbf{beginning of He burning}\\
 \>191-209:\> He burning (first part)\\
 \>210-350:\> blue loop (if any, maximal extension on point 280)\\
 \>351-369:\> He burning (second part)\\
 \>\textbf{370:}\> \textbf{core He exhaustion}\\
 \>371-399:\> C burning\\
 \>\textbf{400:}\> \textbf{last model}
\end{tabbing}

\begin{figure}
\centering
\includegraphics[width=.48\textwidth]{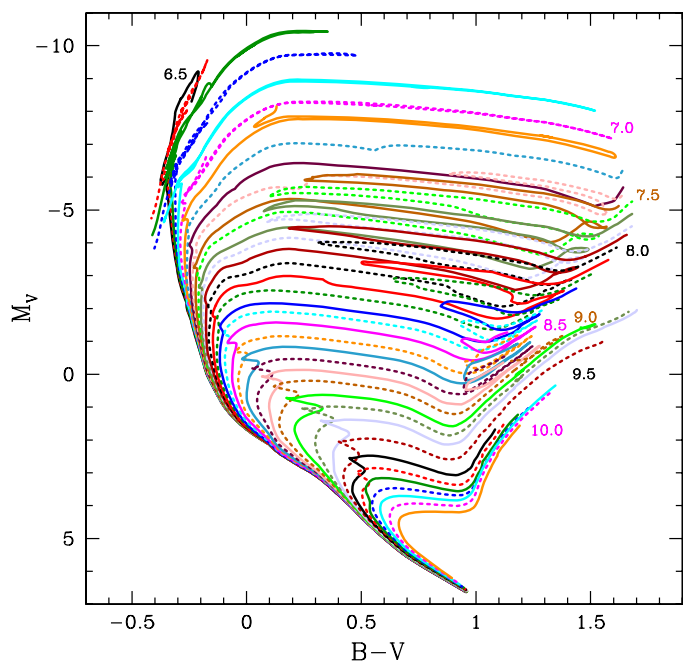}
 \caption{Isochrones from our rotating models. Log ages of 6.5 (top) to 10.1 (bottom) are given, with steps of 0.1 dex.}
      \label{isochr}
\end{figure}
 
For the stars that do not have a certain characteristic, we search for the closest (in mass) stellar model with that feature. The corresponding data point is then determined by the central abundance of the main fuel at that phase in this closest model. For example, the 12 $M_{\sun}$ model does not evolve through the blue loop of a Cepheid. In this model, the point 280 is determined by the central mass fraction of He at the maximal extension of the loop in the 9 $M_{\sun}$ model, which is the closest model with a loop. For the stars that do not go through all the burning phases, the last line of their evolution is repeated up to line 400, so we keep the same size for all the electronic tables.

Isochrones are also provided (see Fig.~\ref{isochr}) on the same webpage. We present the surface properties of the models at the given age, such as $L$, $T_\text{eff}$, $R$, $\log g$, surface abundances, rotational characteristics, as well as colour-magnitude values, such as $M_\text{bol}$, $M_\text{V}$, $U-B$, $B-V$, and $B2-V1$ \citep[conversions from][]{BV81,Flower77,sekifuk00,malagnini86,SchmK82}.

Table \ref{TabGrids} presents the general characteristics of all the models at the end of each burning phase. After the initial mass, initial velocity, and mean velocity on the MS (col. 1 to 3), we give for each burning phase its duration (col. 4, 10, and 16), the total mass (col. 5, 11, and 17), the equatorial velocity (col. 6, 12, and 18), the surface He abundance in mass fraction (col. 7, 13, and 19), and the surface abundances ratios N/C (col. 8, 14, and 20) and N/O (col. 9, 15, and 21). 

When calculating the lifetimes in the central burning stages, we consider the start of the stage as the time when 0.003 in mass fraction of the main burning fuel is burnt. We consider that a burning stage is finished when the main fuel mass fraction drops below $10^{-5}$. 

Table \ref{TabTauM} offers a piecewise linear fit to calculate the MS duration from the mass of the star in the form $\log(\tau_\text{H}) = A \cdot \log(M/M_{\sun}) + B$. The coefficients depend on the mass domain considered. As expected, the lifetime of stars decreases sharply with increasing stellar mass (A$\simeq -3$) in the low mass star regime and the dependence becomes milder (A$\simeq -1$) in the massive star regime. The dispersion around the coefficients is also given in Table~\ref{TabTauM}, and amounts at most to 0.025.

\begin{figure*}
\centering
\includegraphics[width=.98\textwidth]{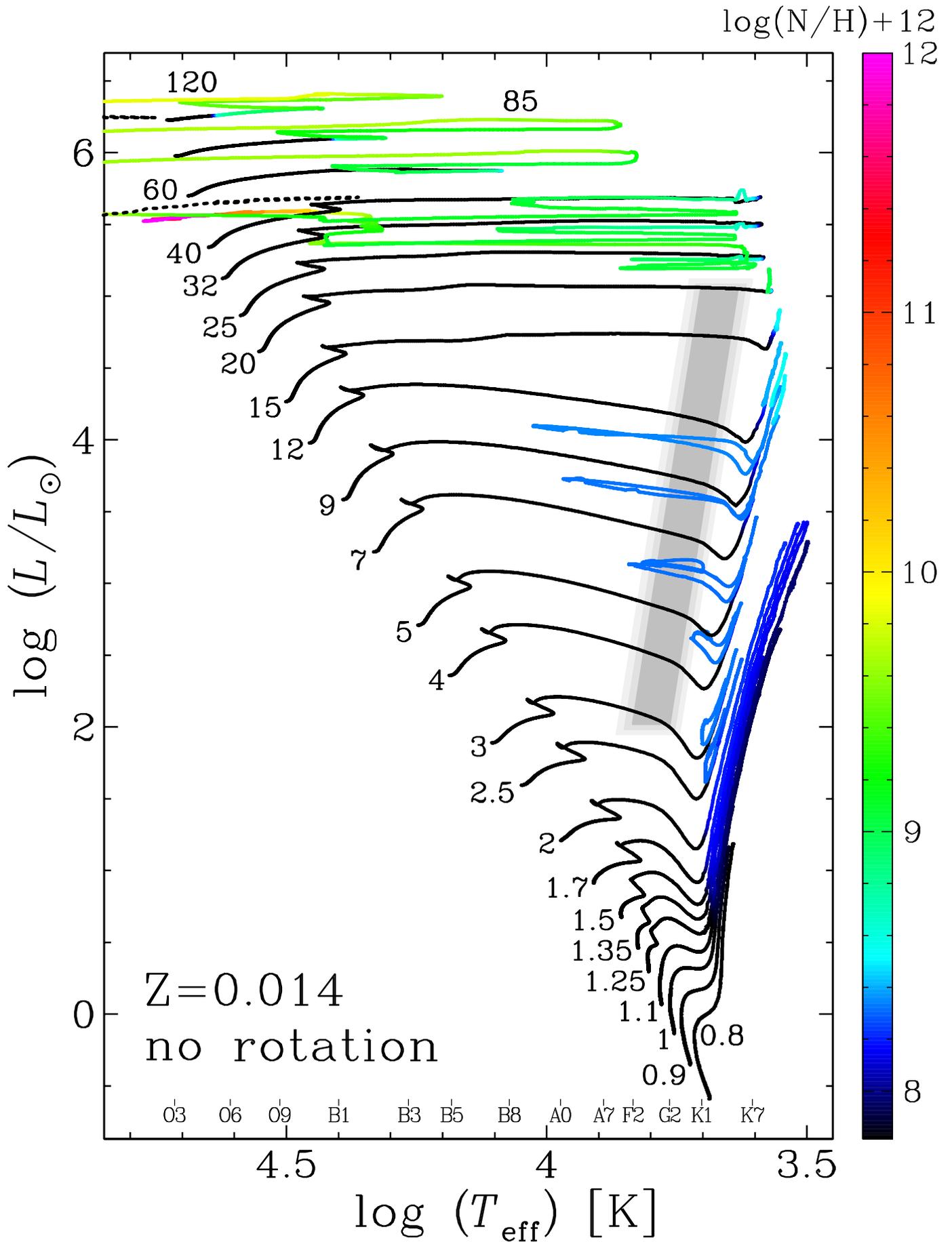}
   \caption{Hertzsprung-Russell diagram for the non-rotating models. The colour scale indicates the surface number abundance of nitrogen on a log scale where the abundance of hydrogen is 12. Once the star has become a WNE type Wolf-Rayet star, the tracks are drawn with black dotted lines. The grey shaded area represents the Cepheid instability strip \citep{tamm03}.}
      \label{FigHRDnorot}
\end{figure*}

\section{Properties of the non-rotating models \label{SecResultsNOROT}}

The present non-rotating tracks can be used for three different purposes: first they can be considered as describing the evolution of slowly rotating stars, {\it i.e.} of stars rotating slowly enough for their outputs to be little affected by rotation. Second they allow comparisons with non-rotating evolutionary tracks that were published previously. Third, since they were computed with exactly the same physics as the rotating tracks, they offer a comparison basis for studying the effects of rotation.

\begin{landscape}
\begin{table}
\caption{Properties of the stellar models at the end of the H-, He-, and C-burning phases.}
\label{TabGrids}
\scalebox{.95}{%
\begin{tabular}{rrr|rrrrrr|rrrrrr|rrrrrr} 
\hline\hline
$M$ & $\upsilon_\text{ini}$ & $\bar{\upsilon}_\text{MS}$ & \multicolumn{6}{c|}{End of H-burning} & \multicolumn{6}{c|}{End of He-burning} & \multicolumn{6}{c}{End of C-burning} \\
  &  &  & $t_\text{H}$ & $M$ & $\upsilon_\text{surf}$ & Y$_\text{surf}$ & N/C & N/O & $t_\text{He}$ & $M$ & $\upsilon_\text{surf}$ & Y$_\text{surf}$ & N/C & N/O & $t_\text{C}$ & $M$ & $\upsilon_\text{surf}$ & Y$_\text{surf}$ & N/C & N/O \\
  $M_{\sun}$ & \multicolumn{2}{c|}{km s$^{-1}$} & Myr & $M_{\sun}$ & km s$^{-1}$ & \multicolumn{3}{c|}{mass fract.} & Myr & $M_{\sun}$ & km s$^{-1}$ & \multicolumn{3}{c|}{mass fract.} & kyr & $M_{\sun}$ & km s$^{-1}$ & \multicolumn{3}{c}{mass fract.}\\
\hline
  120 & 0 & 0 & 2.672 & 63.625 & -- & 0.783 & 94.89 & 77.30 & 0.308 & 30.974 & -- & 0.241 & 0.00 & 0.00 & 0.008 & 30.911 & -- & 0.238 & 0.00 & 0.00 \\
          & 389 & 109 & 3.178 & 34.611 & 1 & 0.974 & 70.15 & 92.07 & 0.350 & 19.045 & 3 & 0.264 & 0.00 & 0.00 & 0.030 & 19.044 & 0 & 0.264 & 0.00 & 0.00 \\
    85 & 0 & 0 & 3.024 & 49.232 & -- & 0.625 & 113.01 & 69.93 & 0.348 & 18.711 & -- & 0.250 & 0.00 & 0.00 & 0.031 & 18.648 & -- & 0.247 & 0.00 & 0.00 \\
          & 368 & 124 & 3.715 & 49.272 & 4 & 0.926 & 81.20 & 84.65 & 0.319 & 26.394 & 19 & 0.264 & 0.00 & 0.00 & 0.013 & 26.393 & 23 & 0.264 & 0.00 & 0.00 \\
    60 & 0 & 0 & 3.530 & 36.261 & -- & 0.493 & 145.35 & 60.81 & 0.397 & 12.565 & -- & 0.279 & 0.00 & 0.00 & 0.115 & 12.495 & -- & 0.273 & 0.00 & 0.00 \\
          & 346 & 138 & 4.465 & 38.536 & 5 & 0.820 & 78.76 & 49.35 & 0.355 & 17.981 & 48 & 0.277 & 0.00 & 0.00 & 0.037 & 17.981 & 22 & 0.277 & 0.00 & 0.00 \\
    40 & 0 & 0 & 4.439 & 36.509 & -- & 0.266 & 0.29 & 0.12 & 0.478 & 13.020 & -- & 0.986 & 63.68 & 87.48 & 0.066 & 12.821 & -- & 0.986 & 33.14 & 28.77 \\
          & 314 & 155 & 5.698 & 32.005 & 20 & 0.554 & 12.73 & 4.31 & 0.424 & 12.334 & 72 & 0.278 & 0.00 & 0.00 & 0.123 & 12.332 & 55 & 0.278 & 0.00 & 0.00 \\
    32 & 0 & 0 & 5.207 & 30.107 & -- & 0.266 & 0.29 & 0.12 & 0.543 & 11.151 & -- & 0.647 & 132.66 & 41.13 & 0.186 & 10.922 & -- & 0.986 & 56.82 & 77.97 \\
          & 306 & 187 & 6.639 & 28.195 & 13 & 0.441 & 5.84 & 1.81 & 0.521 & 10.127 & 46 & 0.280 & 0.00 & 0.01 & 0.213 & 10.125 & 0 & 0.280 & 0.00 & 0.00 \\
    25 & 0 & 0 & 6.310 & 24.180 & -- & 0.266 & 0.29 & 0.12 & 0.692 & 8.757 & -- & 0.537 & 196.18 & 6.69 & 0.433 & 8.289 & -- & 0.831 & 119.91 & 72.30 \\
          & 295 & 209 & 7.902 & 23.604 & 81 & 0.341 & 3.13 & 0.86 & 0.618 & 9.700 & 0 & 0.917 & 102.03 & 23.61 & 0.269 & 9.690 & 0 & 0.926 & 100.50 & 25.12 \\
    20 & 0 & 0 & 7.740 & 19.672 & -- & 0.266 & 0.29 & 0.12 & 0.874 & 9.024 & -- & 0.503 & 59.43 & 3.93 & 1.219 & 8.635 & -- & 0.511 & 81.52 & 4.39 \\
          & 274 & 217 & 9.506 & 19.497 & 176 & 0.300 & 2.42 & 0.60 & 0.846 & 7.191 & 0 & 0.737 & 127.94 & 14.62 & 0.866 & 7.179 & 0 & 0.748 & 127.30 & 15.88 \\
    15 & 0 & 0 & 11.015 & 14.812 & -- & 0.266 & 0.29 & 0.12 & 1.315 & 13.340 & -- & 0.307 & 2.12 & 0.53 & 4.786 & 13.249 & -- & 0.339 & 3.07 & 0.74 \\
          & 271 & 201 & 13.447 & 14.701 & 160 & 0.289 & 2.34 & 0.53 & 1.451 & 11.079 & 0 & 0.427 & 9.40 & 1.33 & 1.374 & 11.071 & 0 & 0.429 & 9.52 & 1.35 \\
    12 & 0 & 0 & 15.330 & 11.940 & -- & 0.266 & 0.29 & 0.12 & 2.124 & 11.364 & -- & 0.301 & 1.84 & 0.49 & 7.927 & 11.308 & -- & 0.307 & 1.98 & 0.53 \\
          & 262 & 198 & 18.369 & 11.907 & 192 & 0.279 & 1.89 & 0.45 & 2.098 & 10.234 & 0 & 0.349 & 6.22 & 0.91 & 3.567 & 10.224 & 0 & 0.350 & 6.27 & 0.92 \\
       9 & 0 & 0 & 26.261 & 8.993 & -- & 0.266 & 0.29 & 0.12 & 3.492 & 8.802 & -- & 0.282 & 1.61 &0.42 & -- & -- & -- & -- & -- & -- \\
          &  248  & 188 & 31.211 & 8.988 & 198 & 0.274 & 1.40 & 0.37 & 3.762 & 8.521 & 4 & 0.337 & 6.09 & 0.86 & 3.143 & 8.517 & 3 & 0.337 & 6.09 & 0.86 \\
\cline{16-21}
       7 & 0 & 0 & 41.721 & 6.999 & -- & 0.266 & 0.29 & 0.12 & 6.919 & 6.915 & -- & 0.279 & 1.52 & 0.40 & \\
          & 235 & 178 & 50.989 & 6.999 & 184 & 0.270 & 0.97 & 0.29 & 6.903 & 6.892 & 2 & 0.322 & 4.68 & 0.76 & \\
       5 & 0 & 0 & 88.193 & 5.000 & -- & 0.266 & 0.29 & 0.12 & 19.380 & 4.959 & -- & 0.279 & 1.48 & 0.39 & \\
          & 219 & 166 & 109.207 & 5.000 & 164 & 0.268 & 0.64 & 0.21 & 17.501 & 4.946 & 3 & 0.318 & 3.79 & 0.69 &  \\
       4 & 0 & 0 & 152.082 & 4.000 & -- & 0.266 & 0.29 & 0.12 & 37.750 & 3.975 & -- & 0.282 & 1.50 & 0.39 \\
          & 197 & 159 & 189.417 & 4.000 & 153 & 0.267 & 0.52 & 0.18 & 33.491 & 3.967 & 4 & 0.319 & 3.38 & 0.66 \\
       3 & 0 & 0 & 320.585 & 3.000 & -- & 0.266 & 0.29 & 0.12 & 117.109 & 2.988 & -- & 0.287 & 1.53 & 0.39 \\
          & 195 & 147 & 405.366 & 3.000 & 136 & 0.267 & 0.43 & 0.16 & 95.566 & 2.984 & 5 & 0.323 & 3.11 &0.63 \\
       2.5 & 0 & 0 & 537.935 & 2.500 & -- & 0.266 & 0.29 & 0.12 & 236.759 & 2.487 & -- & 0.285 & 1.44 & 0.37 \\
          & 187 & 141 & 673.098 & 2.500 & 128 & 0.267 & 0.40 & 0.15 & 181.461 & 2.486 & 5 & 0.322 & 2.92 & 0.59 \\
\cline{10-15}
       2 & 0 & 0 & 1008.831 & 2.000 & -- & 0.266 & 0.29 & 0.12  \\
          & 186 & 137 & 1289.272 & 2.000 & 124 & 0.267 & 0.38 & 0.14 \\
       1.7 & 0 & 0 & 1633.828 & 1.700 & -- & 0.266 & 0.29 & 0.12  \\
          & 176 & 133 & 2098.732 & 1.700 & 120 & 0.268 & 0.35 & 0.13 \\
       1.5 & 0 & 0 & 2241.796 & 1.500 & -- & 0.266 & 0.29 & 0.12  \\
              & 150 & 10 & 2725.402 & 1.500 & 8 & 0.270 & 0.70 & 0.21 \\
       1.35 & 0 & 0 & 3221.584 & 1.350 & -- & 0.266 & 0.29 & 0.12 \\
                & 100 & 7 & 3676.430 & 1.350 & 6 & 0.269 & 0.45 & 0.16 \\
       1.25 & 0 & 0 & 4352.893 & 1.250 & -- & 0.266 & 0.29 & 0.12 \\
                & 100 & 6 & 4680.580 & 1.250 & 4 & 0.270 & 0.43 & 0.15 \\
       1.1 & 0 & 0 & 5464.714 & 1.100 & -- & 0.266 & 0.29 & 0.11 \\
              & 50 & 3 & 5591.467 & 1.100 & 2 & 0.256 & 0.30 & 0.12 \\
       1 & 0 & 0 & 8540.320 & 1.000 & -- & 0.212 & 0.29 & 0.11 \\
          & 50 & 2 & 8788.403 & 1.000 & 2 & 0.253 & 0.30 & 0.12 \\
       0.9 & 0 & 0 & 13461.100 & 0.900 & -- & 0.206 & 0.29 & 0.11 \\
              & 50 & 2 & 13954.450 & 0.900 & 1 & 0.246 & 0.30 & 0.12 \\
       0.8 & 0 & 0 & 21552.716 & 0.800 & -- & 0.195 & 0.29 & 0.11 \\
              & 50 & 1 & 22448.550 & 0.800 & 1 & 0.235 & 0.29 & 0.12 \\
  \cline{1-9}
\end{tabular}}
\end{table}
\end{landscape}

The tracks in the HR diagram for the non-rotating models are presented in Fig.~\ref{FigHRDnorot}. For the morphology of the tracks, we note the following features (going from the massive to the low-mass stars):
\begin{itemize}
\item Owing to the effect of mass loss and high luminosity and $L/M$ ratio, the MS band widens between 40 and 120 $M_{\sun}$, showing a bump around 60 $M_{\sun}$ extending to low effective temperatures down to $\log(T_{\rm eff})=4.1$. 
\item The maximum luminosity of red supergiants is expected to be around $\log(L/L_{\sun})=5.7$.
\item Stars with initial masses between 25 and 40 $M_{\sun}$ evolve back to the blue after a red supergiant phase.
\item Extended blue loops, crossing the Cepheid instability strip, occur for masses between 5 and 9 $M_{\sun}$. 
\item The tip of the red giant branch, {\it i.e.} the luminosity at which the He-flash occurs is around $\log(L/L_{\sun}) = 3.4$ for stars in the mass range between 1.35 and 1.7 $M_{\sun}$. 
\end{itemize}

In Figure \ref{FigHRDnorot}, the colour code indicates the ratio of the nitrogen to the hydrogen number abundances at the surface of the stars, on a logarithmic scale (the initial value is $\log(\text{N/H})+12 = 7.78$). No changes in the surface abundances are expected during the MS phase for stars less massive than 60 $M_{\sun}$. Above that mass limit, the stellar winds are sufficiently strong to uncover nuclear-processed layers already during the MS phase. We note that when the hydrogen surface abundance is between 0.1 and 0.4 (in mass fraction), {\it i.e.} when the star is in the first part of the WNL phase, we have $\log(\text{N/H})+12 \simeq 9.2 - 9.8$ (green portion of the tracks). In those parts, the N/C ratios (in number) are typical of the CNO equilibrium and between 70 and 130 (the initial value is N/C = 0.25). 

For masses below about 40 $M_{\sun}$, the changes in the surface abundances occur after the star has passed through the red supergiant  or red giant stage. The 20 $M_{\sun}$ track is on the verge of evolving back to the blue. Its maximum N/C ratio is already around 70 and its maximum log (N/H)+12 is equal to 9. For masses equal to 15 $M_{\sun}$ and below, the maximum N/C and log (N/H)+12 values are below 3 and 8.6, respectively.

Since the \citet{schaller92} grids, many physical inputs have changed. The initial abundances have been revised, the opacities and reaction rates have been updated, the mass loss prescriptions are not the same, and the overshoot is different. For all these reasons, the comparison of the 92' results with the non-rotating models of the present work reveals many differences.

\begin{figure}
\centering
\includegraphics[width=.48\textwidth]{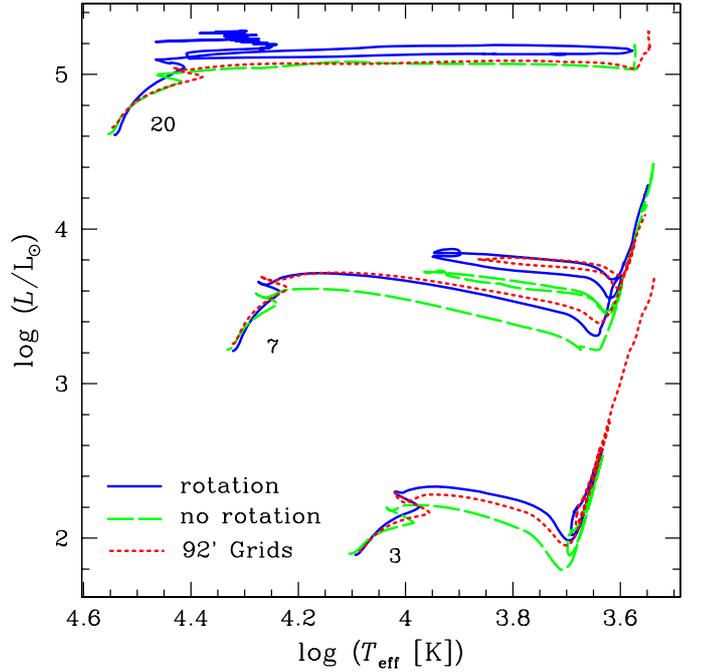}
   \caption{HR diagram of the 3, 7, and 20 $M_{\sun}$ models. Comparison between the present work rotating models (solid blue line), non-rotating models (dashed green line), and the 92' grids models (dotted red line).}
      \label{FigHRD3720}
\end{figure}

Figure~\ref{FigHRD3720} compares the present non-rotating tracks and our previous 92's solar metallicity grid. The most striking difference (compare the long-dashed green curve with the dotted red curve) is the extension of the MS band which is reduced in the present grid as a consequence of the smaller core overshoot (0.10 in the present grid instead of 0.20 in the 92's one). The blue loop for the present 7 $M_{\sun}$ model is more extended and of lower luminosity, as expected for a smaller core overshoot and lower initial metallicity. The mass range going through a Cepheid loop  has also changed: in the 92' grids, the 7 to 12 $M_{\sun}$ models pass through the Cepheid instability strip, while in the new grid the limits are shifted to 5 to 9 $M_{\sun}$.

\begin{figure}
\centering
\includegraphics[width=.48\textwidth]{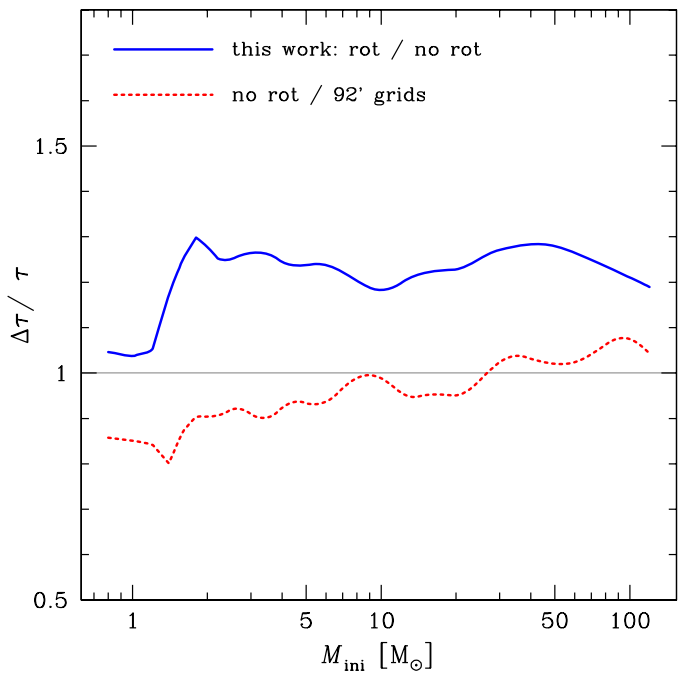}
   \caption{Difference in the MS lifetime duration. Comparison between the present work rotating and non-rotating models (solid blue line), and comparison between the present work's non-rotating  and the 92' grids models (dotted red line).}
      \label{Figtau}
\end{figure}

As shown in Fig.~\ref{Figtau} (dotted red line), the duration of the MS is shorter in the new models, by about 15\% for the lower masses, and 5\% for the intermediate masses. Only the most massive models ($M> 30\, M_{\sun}$) have a longer MS duration by about 4\%, probably owing to the difference in the mass-loss rates. In contrast, the duration of the central He-fusion phases is longer in the new models, by 33-55\% for the intermediate masses and 10-17\% for the massive ones up to 40 $M_{\sun}$. At yet higher masses, the He-burning duration is shorter by 7-26\%. These differences are mainly due to changes in both the overshooting parameter (for masses below 40 $M_{\sun}$) and the mass-loss rates (for masses above 40 $M_{\sun}$).

\begin{figure}
\centering
\includegraphics[width=.48\textwidth]{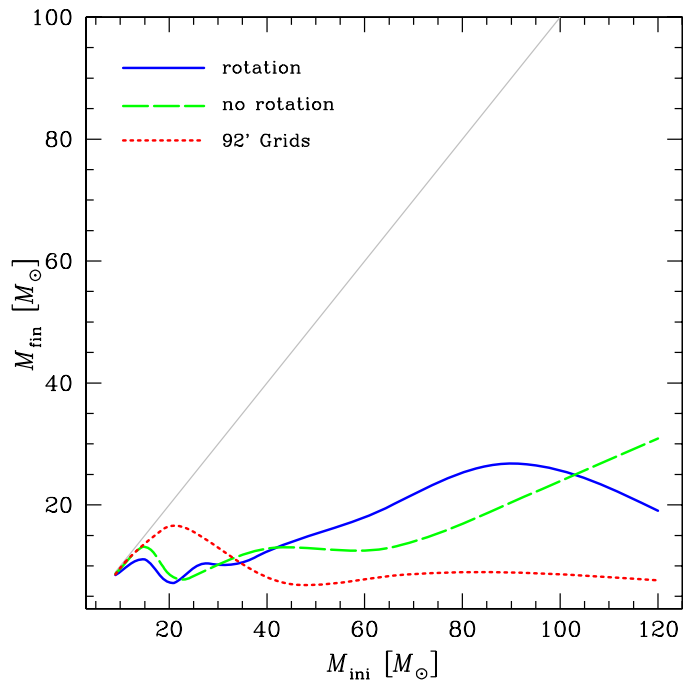}
   \caption{Final mass vs initial mass for models from 9 to 120 $M_{\sun}$. Comparison between the present work's non-rotating models (dashed green line) and rotating models (solid blue line), and the mass loss rates of the 92' grids models computed with the "normal" rates (dotted red line). The grey line corresponds to a hypothetical case without mass loss ($M_\text{fin} = M_\text{ini}$).}
      \label{FigMfin}
\end{figure}

Figure \ref{FigMfin} presents the final masses obtained for our models and the ones given in \citet{schaller92}. We see that for the lower mass range, the present models deviate from the slope one relation (grey line) at lower mass than the 92's models, typically around 15 $M_{\sun}$ instead of 25 $M_{\sun}$. This is a consequence of the prescription adopted for the mass-loss rate during the red supergiant stage (cf. Sect.~\ref{SubSecMdotEdd}) which leads to stronger mass loss during that phase in the present models. 

In the upper mass range (above 35 $M_{\sun}$), the present non-rotating models have higher masses at the end of the C-burning phase than the masses we obtained in 92. This is a consequence of the lower mass-loss rates used here during the Wolf-Rayet phases.

\begin{table}
\centering
\caption{Coefficients of a piecewise linear fit to calculate the MS duration from the mass of the star.}
\label{TabTauM}
\begin{tabular}{c l |c c c}
\hline\hline
 & &\multicolumn{2}{c}{$\log(\tau_\text{H}\,[\text{yr}]) = A \cdot \log(M/M_{\sun}) + B$} & standard\\
    mass range & & A & B & deviation \\
\hline
    1.25 - 3.0 & no rot. & $-$2.926 & 9.892 & 0.016 \\
    &       rot. & $-$2.776 & 9.938 & 0.012 \\
    & & & &  \\
    3 - 7    & no rot.  & $-$2.405 & 9.641 & 0.013 \\
    &       rot.  & $-$2.444 & 9.761 & 0.013 \\
    & & & &  \\
    7 - 15  & no rot.  & $-$1.765 & 9.105 & 0.010 \\
    &       rot.  & $-$1.763 & 9.186 & 0.015 \\
    & & & &  \\
    15 - 60  & no rot.  & $-$0.808 & 7.954 & 0.025 \\
    &       rot.  &  $-$0.775 & 8.004 & 0.022 \\
\hline
\end{tabular}
\end{table}

\section{Properties of the rotating models \label{SecResultsROT}}

\subsection{Main sequence width \label{SubSecMSWidth}}

\begin{figure}
\includegraphics[width=.48\textwidth]{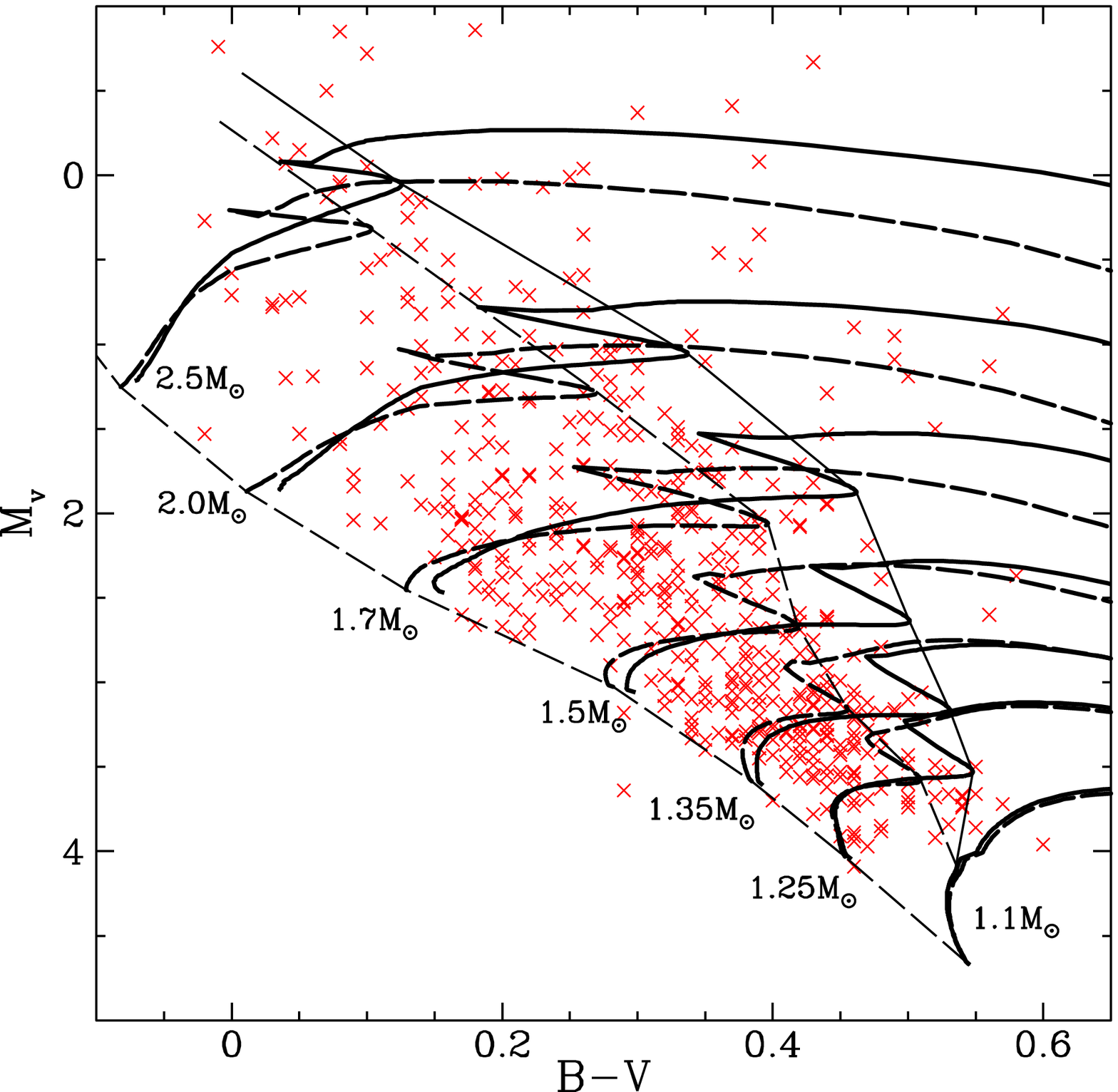}
   \caption{Evolutionary tracks for non-rotating (dashed lines) and rotating (solid lines) models in the colour-magnitude diagram. The ZAMS and the minimal MS $T_\text{eff}$ are indicated as well as the initial mass considered. The crosses are observed MS intermediate mass stars from Tables 1 and 2 of \citet{wolff97}.}
      \label{Wolff2}
\end{figure}

In Fig.~\ref{Wolff2}, we superimpose the evolutionary tracks of the present paper with rotation (solid lines) and without rotation (dashed lines) on the data points for the intermediate-mass main-sequence stars discussed in \citet{wolff97}. We see that for masses above about 1.7 $M_{\sun}$, non-rotating tracks computed with an overshoot parameter $d_\text{over}/H_P=0.10$ predict a too narrow MS band, while the tracks computed with an average rotation and this same overshoot parameter closely fit the observed MS band. Thus, in this mass range, the effects of a moderate overshoot and rotational mixing closely fit the observations.

For masses between 1.25 and 1.5 $M_{\sun}$, the situation is much less clear: the non-rotating MS tracks are in general too narrow and the rotating ones too wide. This mass domain just corresponds to the upper mass range where magnetic braking comes into play. We note that with our present prescription for magnetic braking, calibrated for the solar case, the mean velocity during the MS phase is between 6 and 10 km s$^{-1}$. This is smaller than the average observed rotations for this mass range, which are between 10 and 40 km s$^{-1}$ for field stars and between 20 and 130 km s$^{-1}$ for cluster stars according to \citet{wolff97}. Hence, it may be that the present models, in this mass range, overestimate the magnetic braking, as we will investigate in a future paper. Until then, we note that rotating and non-rotating models encompass the observed MS widths in this mass range.

\citet{meulenaer2010}, using asteroseismology, found that $\alpha$ Cen A (a G2V star with a mass of 1.105 $\pm$ 0.007 $M_{\sun}$) has observed asteroseismic properties that can be more accurately reproduced by models with no convective core confirming the choice we made to account for no overshoot in models below 1.25 $M_{\sun}$.

With the CoRoT satellite, \citet{briquet2011} observed the O9V star HD 46202 (member of the young open cluster NGC 2244). The mass of this star is around 20 $M_{\sun}$, its $\upsilon \sin i$ is $\sim$25 km $s^{-1}$, and it is on the main sequence (at an age estimated to be around 2 Myr). If this star were a truly slow rotator (namely if the low $\upsilon \sin i$ were not due to a $\sin i$-effect and the star were throughout its lifetime a slow rotator), then the extension of the core would be entirely due to overshooting, with no or an extremely small contribution from rotational mixing. Very interestingly, their best-fit model gives a core overshooting parameter $d_\text{over}/H_P=0.10\pm0.05$, quite in agreement with the overshoot used in the present models. In contrast, \citet{lovekin2010}, studying $\theta$ Ophiuchi, a MS $\sim$9.5 $M_{\sun}$ star with a  $\upsilon \sin i$ of 30 km $s^{-1}$, obtain an overshooting parameter equal to 0.28$\pm$0.05. If this star were a truly slow rotator, and thus the extension of the core due mainly to the overshooting process, the result obtained in this work would be at odds with our choice of overshooting parameter. This illustrates that despite many decades of discussion about the overshooting parameter, its precise value remains uncertain.

\begin{figure*}
\centering
\includegraphics[width=.98\textwidth]{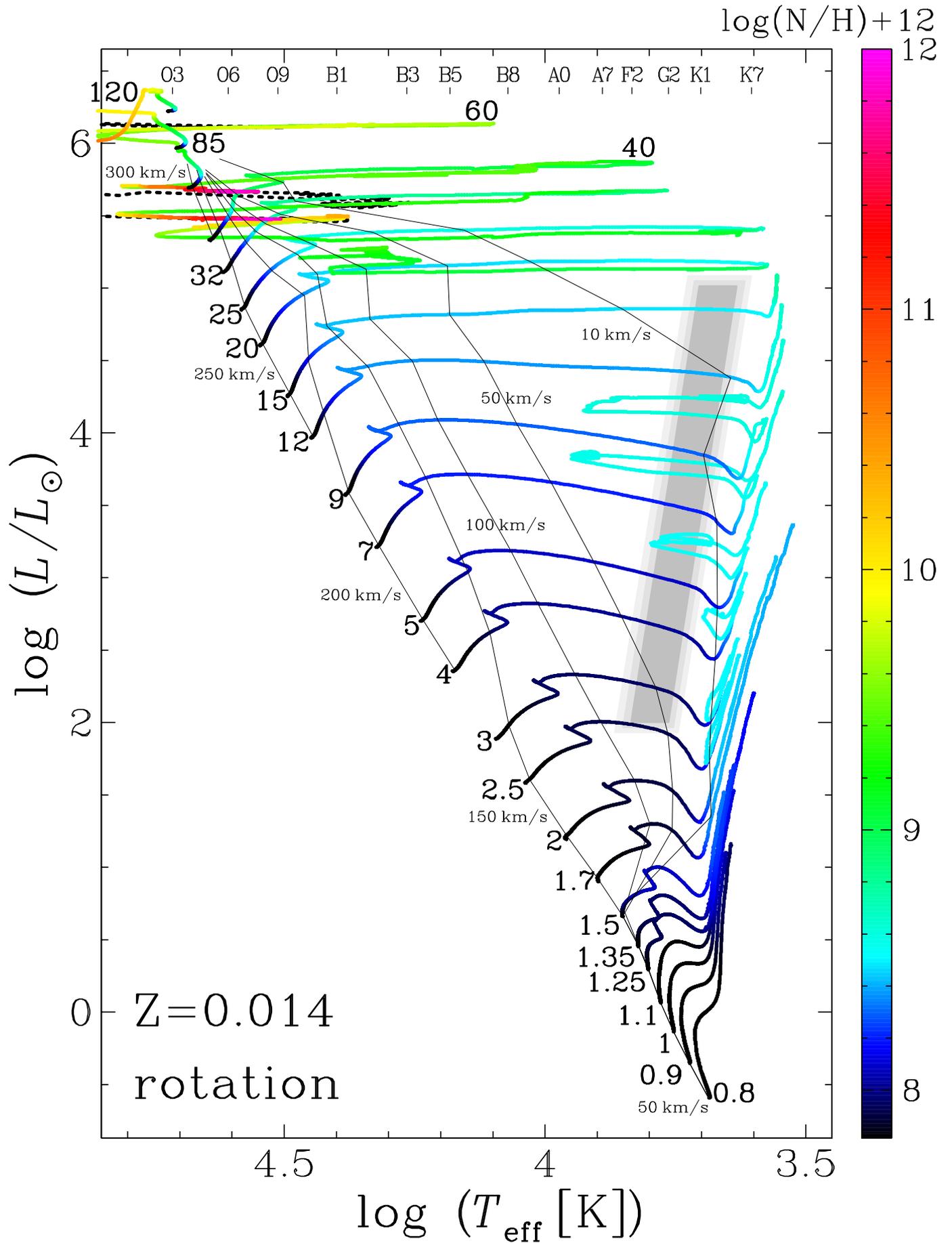}
   \caption{Same as Fig.~\ref{FigHRDnorot} but for the rotating models. Lines of iso-velocities are drawn through the diagram.}
      \label{FigHRDrot}
\end{figure*}

\subsection{Hertzsprung-Russell diagram and lifetimes}

The tracks in the HR diagram for all the rotating models are presented in Fig.~\ref{FigHRDrot}, with lines of iso-velocities drawn across the diagram. Compared to Fig.~\ref{FigHRDnorot}, we note the following differences:
\begin{itemize}
\item There is no longer any widening of the MS band occurring around the 60 $M_{\sun}$ stellar model as in the non-rotating grid. This is a consequence of rotational mixing, which prevents any significant redward evolution during the MS phase. \textit{This shows that in this upper mass range, the MS band width is very sensitive to rotational mixing.}
\item The maximum luminosity of red supergiants is in the range $\log(L/L_{\sun}) =  5.2-5.4$, hence is a factor of two inferior to the limit obtained with non-rotating models. This value closely agrees with the upper limit given by \citet{levesque05}.
\item Stars with initial masses between 20 and about 25 $M_{\sun}$ evolve back to the blue after a red supergiant phase. This range is between 25 and 40 $M_{\sun}$ in the non-rotating grid.
\item Extended blue loops, crossing the Cepheid instability strip occur for masses between 5 and 9 $M_{\sun}$ as for the non-rotating tracks. For a given mass, the loop however occurs at a higher luminosity. The blueward extension is similar or slightly shorter.
\item The tip of the red giant branch does not appear to be very different in the rotating models. We must stress here that not all the models have been computed strictly up to the He-flash, preventing us from reaching more definite conclusions. The above assessment relies on the case of the 1.7 $M_{\sun}$ model that was computed up to that stage. 
\item Changes in the surface abundances occur at a much earlier evolutionary stage (see Sect.~\ref{SubSecSurfAbund} below).
\end{itemize}

\begin{figure*}
\centering
\includegraphics[width=.45\textwidth]{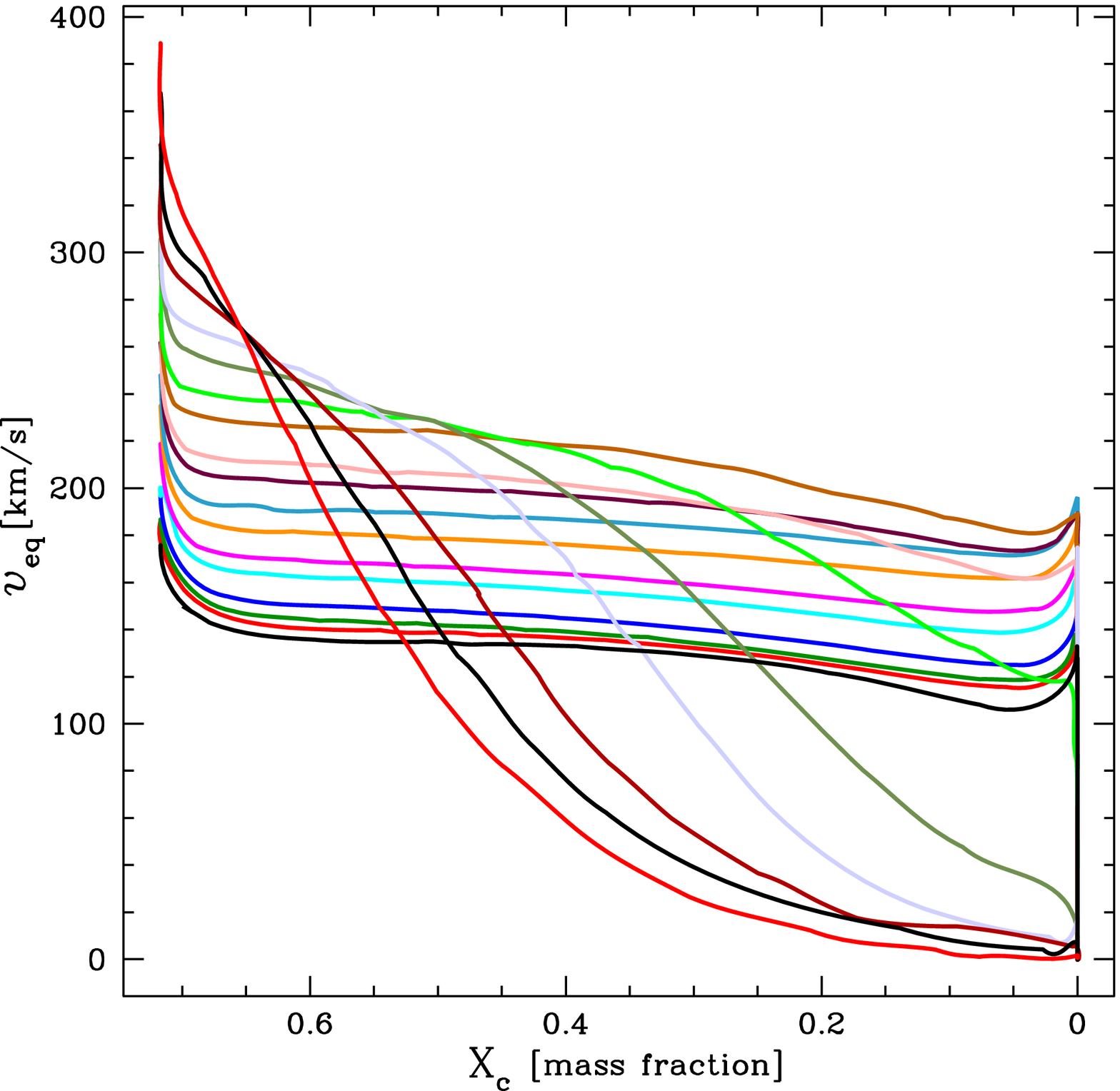}\hspace{.5cm}\includegraphics[width=.45\textwidth]{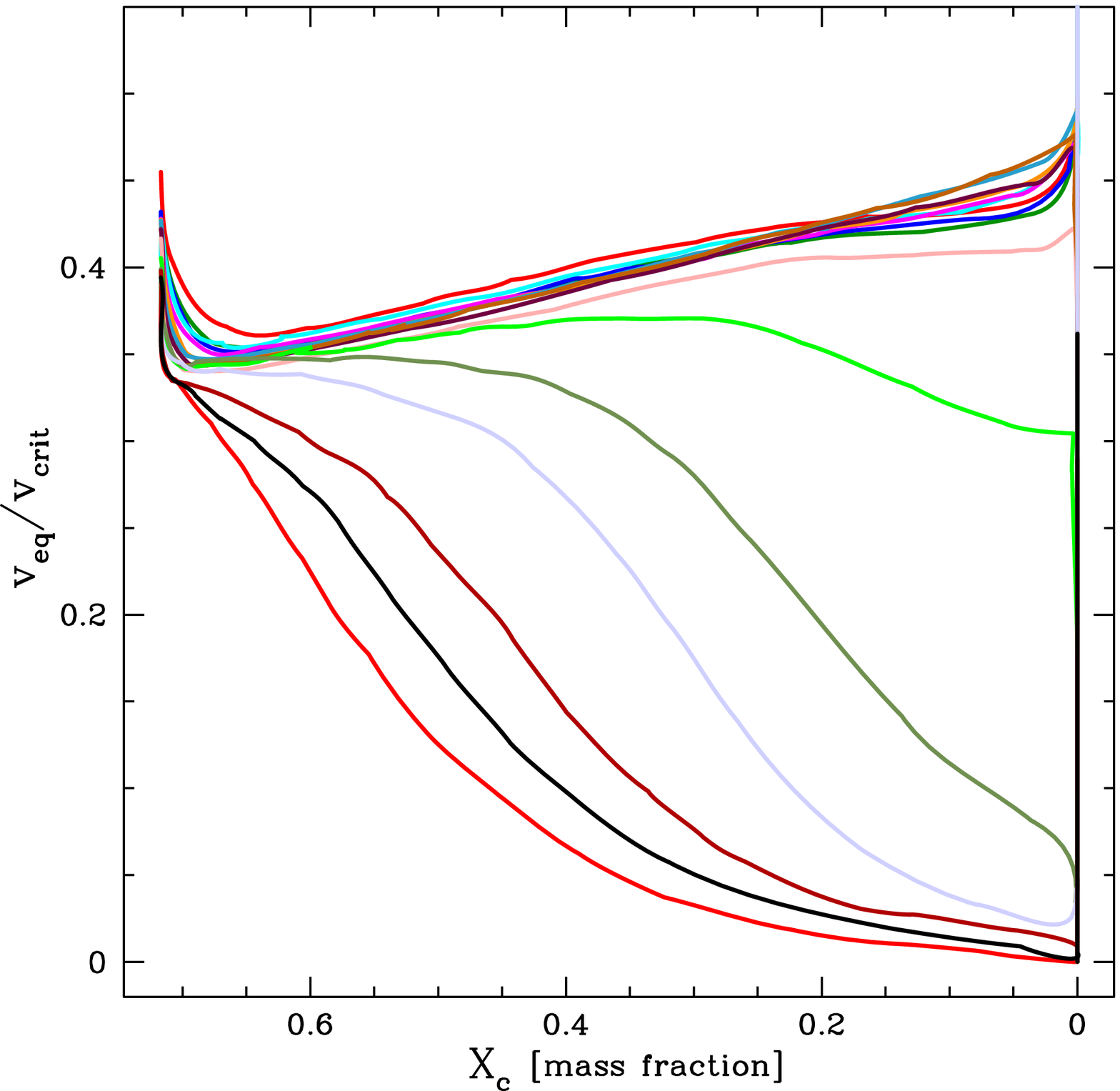}
   \caption{Evolution of rotation on the main sequence as a function of the central mass fraction of hydrogen for all the stellar models from 1.7 to 120 $M_{\sun}$. \textit{Left:} Evolution of the equatorial velocity. The curve for the 1.7 $M_{\sun}$ is the bottom line at the beginning of the evolution (X$_\text{c}=0.72$), and that for the 120 $M_{\sun}$ the upper line (increasing equatorial velocity for increasing mass). \textit{Right:} Evolution of the $\upsilon_\text{eq}/\upsilon_\text{crit}$ ratio. The curve for the 1.7 $M_{\sun}$ is the upper line during the MS, and that for the 120 $M_{\sun}$ the bottom line during the MS (decreasing velocity ratio for increasing mass).}
      \label{FigRotEvol}
\end{figure*}

Figure~\ref{FigHRD3720} directly compares the rotating and non-rotating tracks for the 3, 7, and 20 $M_{\sun}$ models. We see that the present rotating tracks end the MS phase at lower effective temperature and higher luminosity than the non-rotating ones as found by previous studies \citep{HL00,mm5}. They cross the Hertzsprung gap at a higher luminosity than their non-rotating counterparts. This shift in luminosity persists for the remainder of the evolution. As a consequence, the Cepheid loop of the 7 $M_{\sun}$ occurs at a luminosity around 0.15 dex higher than in the non-rotating case. We note that this corresponds to an evolutionary mass that is about 15\% lower for a Cepheid at a given luminosity. It has been proposed that mass loss and internal mixing might reconcile the evolutionary and pulsational masses for Cepheids \citep{Keller08,neilson11}: rotation would account for both effects and the masses inferred would thus be closer to the pulsational ones. The rotating 20 $M_{\sun}$ evolves back to the blue in contrast to the non-rotating one, indicating that rotational mixing, together with enhanced mass loss during the red supergiant stage, promote blueward evolution \citep[see also][]{hmm04}.

As shown in Fig.~\ref{Figtau} (blue solid line), rotation generally enhances the time spent on the MS, by about 25\% throughout all the masses above 2 $M_{\sun}$. The increase in the MS lifetime is caused by rotational mixing, which refuels the core in fresh hydrogen. It is striking to see how relatively constant the increase in the MS lifetime appears to be over this whole mass range. This shows that the effect of rotation scales very well with the initial $\upsilon_{\rm ini}/\upsilon_{\rm crit}$ value. The picture is much less clear for the core helium- and carbon-burning lifetimes (not shown in Fig.~\ref{Figtau}). Some models have an extended lifetime, others shorter. The net result depends on many parameters, such as the choice of diffusion coefficients  (which governs the size of the convective core and the mixing from or to the adjacent H- and/or He-burning shell), or the mass loss.

Figure~\ref{FigMfin} compares the final and initial mass relations. For masses lower than 40 $M_{\sun}$, the difference obtained between the rotating and non-rotating models is small. Above that mass, the rotating models up to 100 $M_{\sun}$ lose less mass than the non-rotating one. Only in the very upper mass range do the rotating models end with lower final masses than the non-rotating ones. This is caused by the non-rotating stars spending more time in regions of the HR diagram where the mass loss is stronger. By remaining in the blue part of the HR diagram, the rotating models pass from O-type mass-loss rates directly to WR mass-loss rates, without experiencing the bistability jump, and the red supergiant (RSG) and supra-Eddington mass-loss rates.

\subsection{Evolution of the surface velocities}

Figure \ref{FigRotEvol} presents the evolution of the equatorial velocity (\textit{left}) and the $\upsilon_\text{eq}/\upsilon_\text{crit}$ ratio (\textit{right}), during the MS for all the models. Since we computed models with identical initial $\upsilon_\text{eq}/\upsilon_\text{crit}$ ratio, the initial equatorial velocity varies with the mass considered: higher masses need a larger equatorial velocity to attain a given ratio. In Fig.~\ref{FigRotEvol} (\textit{left}), this is clearly visible: the 1.7 $M_{\sun}$ model draws the bottom black line, starting its MS evolution with $\upsilon_\text{eq}=170$ km s$^{-1}$, while the most massive 120 $M_{\sun}$ model draws the upper red line on the ZAMS, starting its MS evolution with $\upsilon_\text{eq}=389$ km s$^{-1}$. 

We bring our models onto the ZAMS assuming solid body rotation, an assumption that is no longer made after the ZAMS. As soon as we allow for differential rotation, a readjustment of the $\Omega$-profile occurs inside the model, which explains the rapid drop in equatorial velocity at the very beginning of the evolution \citep{diw99}. Once the equilibrium profile is reached, the equatorial velocity evolves under the action of the internal transport mechanisms (which tend to bring angular momentum from the contracting core towards the external layers, spinning up the stellar surface)\footnote{Even though the net transport of angular momentum is towards the surface, approximating the meridional circulation (an advective process) as a diffusive process would lead to even more transport of angular momentum from the core to the surface and to less shear mixing and thus less mixing of elements like nitrogen to the surface.} and the mass loss (which removes angular momentum from the surface, spinning down the star). 

In low- and intermediate-mass stars, the transport mechanisms are quite inefficient, but at the same time, mass loss is either absent or very weak (at least during the MS). Stars thus evolve with a quasi constant $\upsilon_\text{eq}$. This agrees with the results of \citet{wolff07}, who find that the rotation rates for stars in the mass range 6-12 $M_{\sun}$ do not change by more than 0.1 dex over ages between about 1 and 15 Myr.

With increasing mass, the internal transport of angular momentum becomes more efficient, but at the same time the mass loss becomes dominant, so the most massive models ($M\ge32\,M_{\sun}$) quickly spin down. Therefore we see that the time-averaged surface velocity is smaller  (see Table 2) for stars in the mass range between 32 and 120 $M_{\sun}$ (between 100 and 180 km s$^{-1}$), than in the mass range from 9 to 25 $M_{\sun}$ ($\sim$ 200 km s$^{-1}$).

It is interesting to compare the mean rotation velocity over the whole main sequence $\bar{\upsilon}_{MS}$ with the velocity of the stars at the end of the MS phase (see Table 2). We see that for stars with initial masses equal or superior to 32 $M_{\sun}$, the surface velocity at the end of the MS phase is much smaller than $\bar{\upsilon}_{MS}$, reaching a minimum of only a few km s$^{-1}$ and reaching at most a value of 20 km s$^{-1}$. For these stars, we observe what might be referred to as a very strong mass-loss braking mechanism. The situation is qualitatively the same for stars in the mass range between 12 and 25 $M_{\sun}$, although less spectacular since the mass-loss rates are much lower. For the masses below 9 $M_{\sun}$ down to 1.7 $M_{\sun}$, the situation is different: the velocities at the end of the MS phase are nearly equal, and sometimes even slightly superior to $\bar{\upsilon}_{MS}$. This illustrates, as already indicated above, that in the absence of stellar winds, the surface continuously receives angular momentum transported by the meridional currents, and to a smaller extent by shears. If this were not the case, the surface would slow down simply as a result of its inflation and the local conservation of the angular momentum. For masses equal to or below 1.5 $M_{\sun}$, we see that $\bar{\upsilon}_{MS}$ is also very similar to the velocities at the end of the MS phase, although much smaller than the initial velocities. This is the result of the very rapid braking exerted by the magnetic fields at the beginning of the evolution.  

After the MS phase, the surface velocities and their evolution are depicted by the lines of iso-velocities drawn across the diagram in Fig.~\ref{FigHRDrot}. The surface velocity decreases rapidly during the crossing of the Hertzsprung gap for all masses. For all stars between 2 and 12 $M_{\sun}$, the iso-velocity line corresponding to 10 km s$^{-1}$ is nearly vertical with an abscissa in the range $\log(T_{\rm eff}) = 3.65 - 3.7$. When a blue loop occurs, the surface velocity increases, sometimes reaching the critical velocity as in the case of the 9 $M_{\sun}$ model. In this case, however, the proximity to the critical limit occurs over such a short duration that the star undergoes very little mechanical mass losses.

Looking at the velocities obtained at the end of the core He-burning phase in Table~\ref{TabGrids}, we see that the stars finishing their lives as WR stars have surface velocities between only a few km s$^{-1}$ as the 120 $M_{\sun}$ model, up to slightly more than 70 km s$^{-1}$ for the 40 $M_{\sun}$ model. This is in good agreement with the observations of WR stars by \citet{CStL08}, who obtain rotational velocities of between 10 and 60 km s$^{-1}$ from periodic large-scale wind variability.

Figure \ref{FigRotEvol} (\textit{right}) illustrates that the overall evolution of the $\upsilon_\text{eq}/\upsilon_\text{crit}$  ratio looks different from that of $\upsilon_\text{eq}$, at least for the models that do not undergo strong braking by mass loss. This is due to the evolution of $\upsilon_\text{crit}$ itself: in the framework of the Roche model we use here, we have $\upsilon_\text{crit}\propto\sqrt{M/R}$. During the MS, $M$ decreases with the mass loss, and $R$ inflates, so $\upsilon_\text{crit}$ decreases. This explains why models of low- and intermediate-mass that have a quasi constant or slightly decreasing $\upsilon_\text{eq}$ have the $\upsilon_\text{eq}/\upsilon_\text{crit}$ ratio that increases during the MS evolution. A more detailed discussion of this subject can be found in \citet{emmb08}.

\subsection{Evolution of the surface abundances \label{SubSecSurfAbund}}

\begin{figure}
\centering
\includegraphics[width=.48\textwidth]{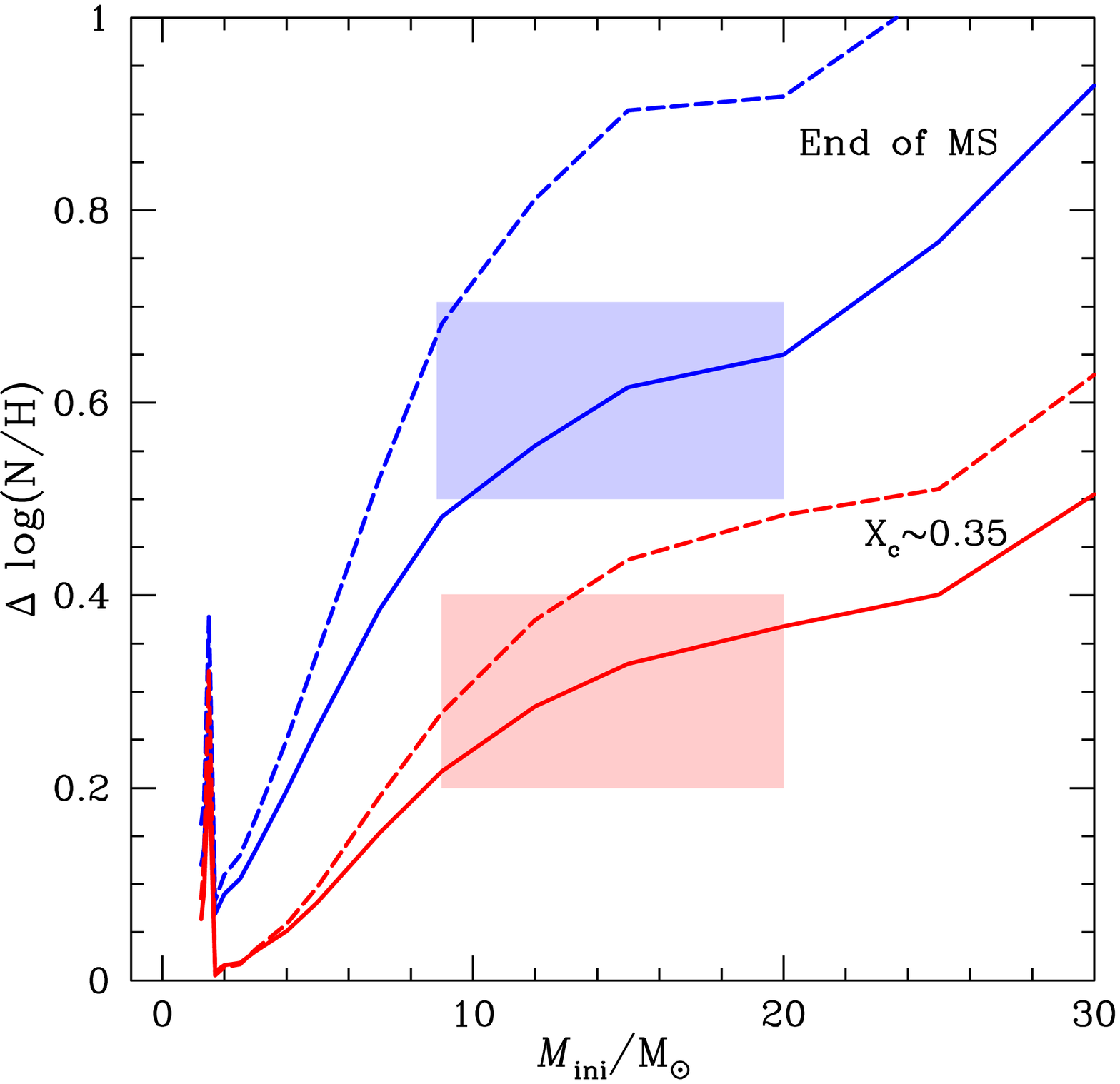}
 \caption{The solid curves show the variation as a function of the initial mass of the N/H ratio at the surface at, respectively, the end (upper solid blue line labelled by ``End of MS") and at the middle of the MS phase (lower solid red line labelled by ``$X_\text{c}\sim0.35$"). The lower hatched zone corresponds to mean values of the N/H excess observed at the surface of Galactic MS B-type stars with initial masses inferior to 20 $M_{\sun}$ \citep{gieslamb92,Kilian92,morel08,hunter09}. The upper one corresponds to the observed range for maximum values. The dashed curves show the variation in the N/C ratio normalised to the value on the ZAMS.}
      \label{FigNHbis}
\end{figure}

\begin{figure*}
\centering
\includegraphics[width=.45\textwidth]{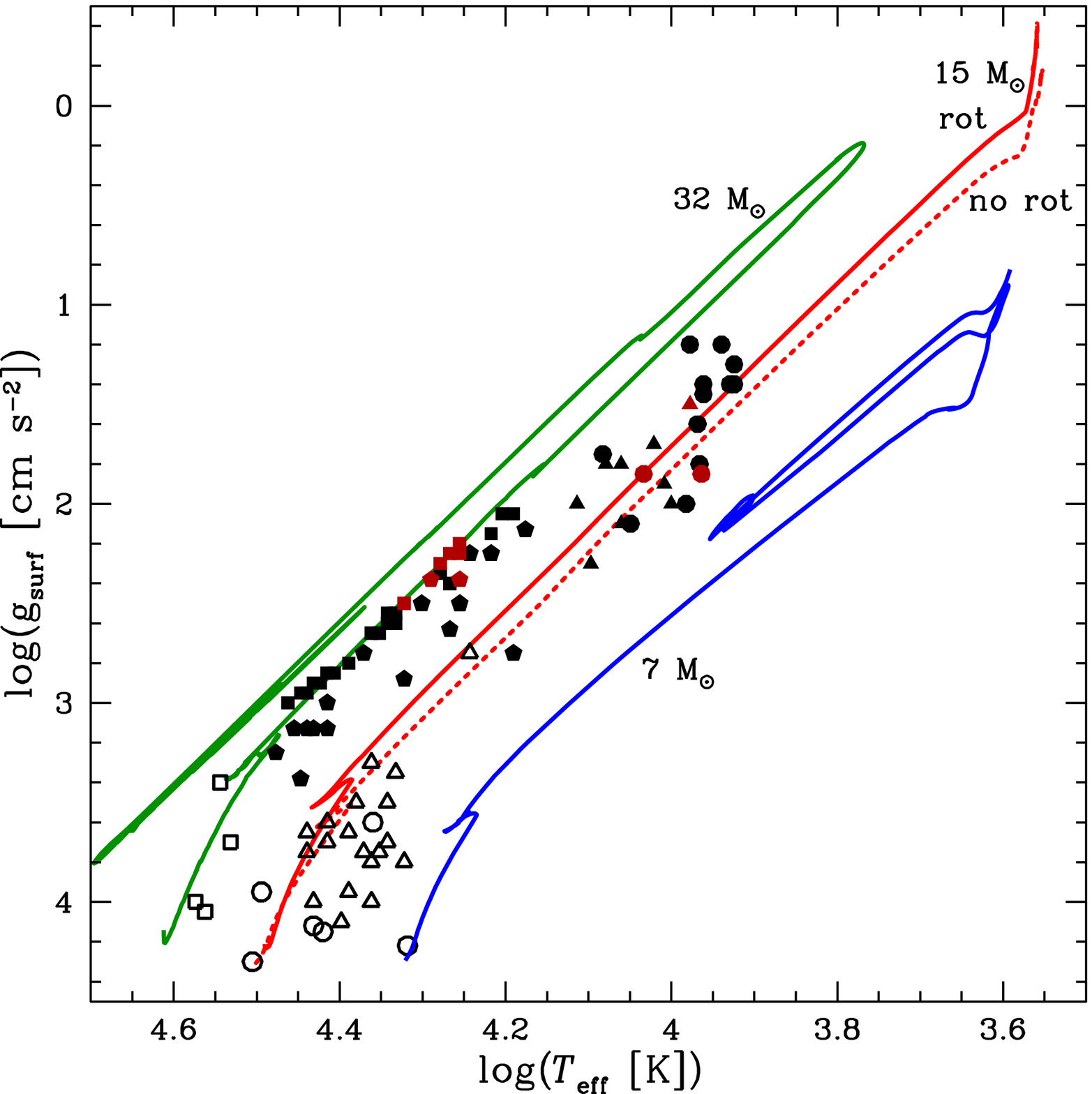}\hspace{.3cm}\includegraphics[width=.45\textwidth]{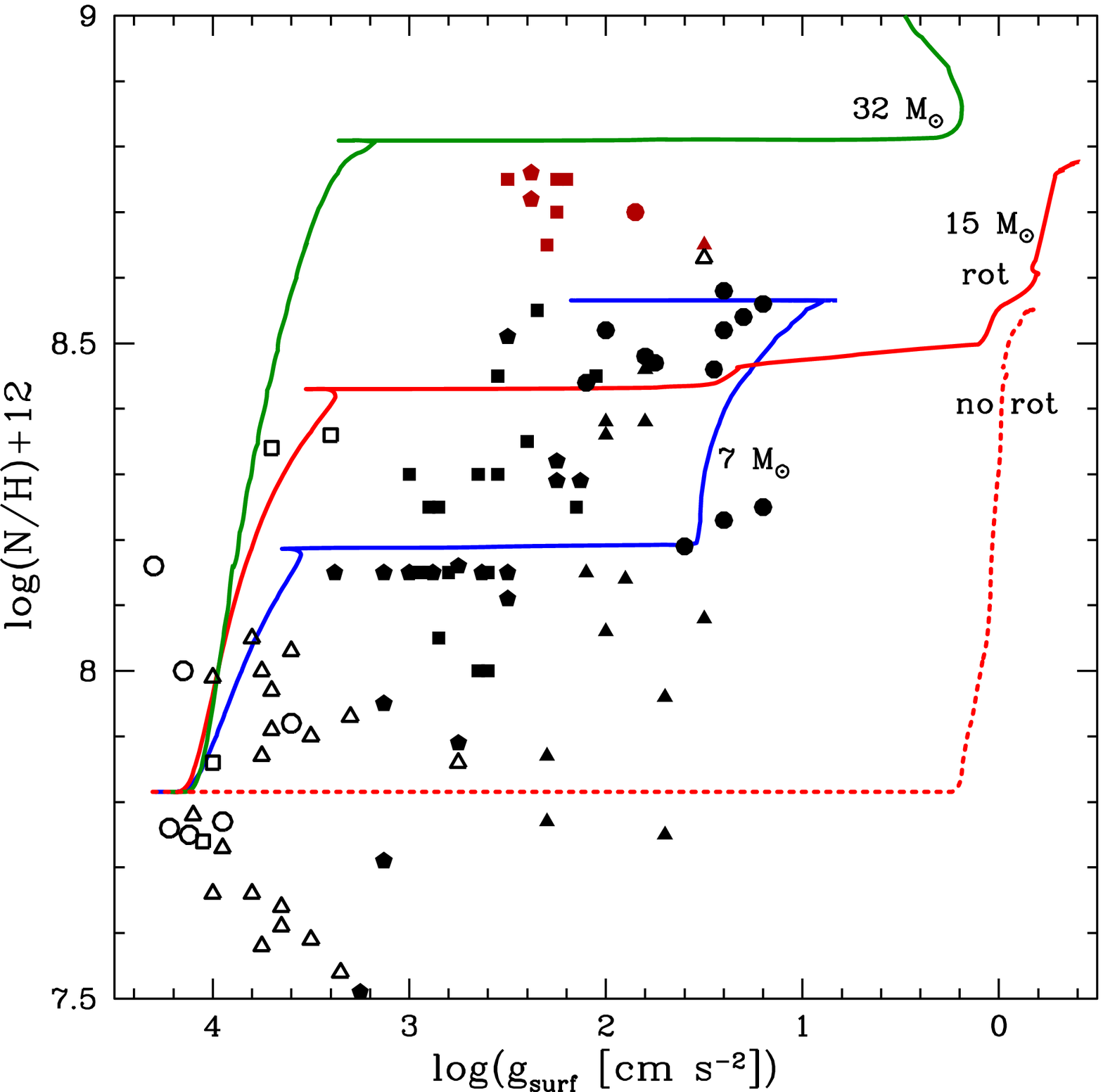}
   \caption{$\log(g) - \log(T_\text{eff})$ diagram (\textit{left}) and evolution of the abundance ratio N/H (\textit{right}) for the 7 (blue), 15 (red), and 32 $M_{\sun}$ (green) rotating models. The track of the non-rotating 15 $M_{\sun}$ is indicated with the red dotted line. Observational points from O- and B-type stars are plotted, with empty symbols for luminosity class (LC) V-III (triangles: \citealp{morel08}; squares: \citealp{villamariz02}, \citealp{VH2005}; circles: \citealp{przybilla2010}) and filled symbols for LC I (triangles: \citealp{TTH2000}; squares: \citealp{crowther06}; hexagons: \citealp{searle08}; circles: \citealp{przybilla2010}).}
      \label{FigNHg}
\end{figure*}

The evolution of the surface abundances depends upon both the mixing processes, which modify the composition of the internal stellar layers, and the mass loss, which removes the external layers and uncovers the deeper regions. Rotation, by driving mixing and enhancing mass loss is very efficient in enriching the surface, as has been shown by previous studies \citep{HL00,mm5}. The colour code in Fig.~\ref{FigHRDrot} illustrates this. The variation in the N/H ratio at the surface of various initial mass models at two evolutionary stages during the MS phase is shown in Fig.~\ref{FigNHbis}. Only models with rotation are shown, since, for the mass range considered here, non-rotating models would not show any changes in the surface abundances and their track would be a horizontal line at an ordinate equal to 0, as can also be seen in Fig.~\ref{FigHRDnorot}.

We see that changes in the surface abundances occur as soon as the MS phase for all masses above 1.25 $M_{\sun}$ (see also Table~\ref{TabGrids}). The [N/H] ratio at the middle of the MS is higher by more than a factor of two at the surface of all stars with $M > 13 M_{\sun}$ (see Fig.~\ref{FigNHbis}). If we consider the end of the MS phase, the minimum initial mass for which there is a factor of two enhancement in the [N/H] ratio is less than 6 $M_{\sun}$.  The variations in the [N/C] ratio follow the same trend as the [N/H] ratio. It is simply shifted to higher values since the carbon abundance varies faster than the hydrogen abundance at the surface of the stars.

As already discussed in \citet{maeder09}, the enrichment attained depends on both the initial mass (larger at higher mass) and the age (larger at more advanced stages). We see that this trend continues down to 1.7 $M_{\sun}$. Below that mass, magnetic braking, by enforcing a large gradient in the rotational velocities in the outer layers, boosts the mixing of the chemical species \citep[see also][]{meynet11mag} and produces the peak seen in the low mass range in Fig.~\ref{FigNHbis}. As mentioned in Sect.~\ref{SubSecMSWidth}, the magnetic braking might be overestimated in these low-mass models, since the predicted enrichment is higher than expected in this mass domain. A smaller magnetic braking would probably allow the models to closely fit both the observed surface velocities and enrichments in this mass range. The present models, however,  give the consistent solution obtained when  a solar magnetic braking law is applied to stars with initial masses between 1 and 1.5 $M_\odot$.

Looking at Fig.~\ref{FigNHbis}, we see also that the curves form a kind of plateau in the mass range between 15 $M_{\sun}$ and 20-25 $M_{\sun}$. This plateau is the result of two competing effects: first, as discussed in the review by Maeder \& Meynet (2011, in press), for a given interior differential rotation, mixing becomes more efficient in increasing initial stellar masses; second, when the initial mass increases, meridional currents also become more rapid in the outer layers; they are thus more efficient at eroding the rotational velocity gradients and make rotational mixing less efficient.  Above the plateau, the effects of mass loss in uncovering the inner layers come into play, and the curves then increase more rapidly.

The evolution of the surface abundances after the MS phase in the rotating models is shown in colour scale in Fig.~\ref{FigHRDrot}, and the N/C and N/O ratios are given in Table~\ref{TabGrids}. When the star first crosses the Hertzsprung gap, its surface abundance is the same as the one reached at the end of the MS phase, since that phase is too short to allow any further changes in the surface abundances. As a consequence, slowly or non-rotating MS stars are not expected to show any changes in the surface abundances when they are on their first track towards the red supergiant stage, while those originating from moderately rotating progenitors on the MS band, would differ away from the initial abundances.

During the red supergiant phases, some additional changes in the surface abundances occur, resulting from the dredge-up and additional rotational mixing. Thus, the stars originating from moderately rapid rotators at that stage undergo stronger enrichments than those originating from either slowly or non-rotating stars. Typically, at the end of the core He-burning phase, the N/C ratios are of the order of 2 in our non-rotating models with initial masses between 2.5 and 15 $M_{\sun}$, while they are between 3 and 10 in the corresponding rotating models, those of higher masses presenting the higher N/C values. The values for the N/O ratio are between 0.4 and 0.5 for the non-rotating models and between 0.6 and 1.3 for the rotating ones.

As a consequence of the high mass-loss rates experienced by the stars between 20 and 32 $M_{\sun}$ during the red supergiant phase, strong variations in the N/C and N/O ratios are obtained. Values as high as those obtained at CNO-equilibrium are reached or in contrast, very low values owing to the appearance of He-burning products at the surface (because N is destroyed during core He burning).

Above 40 $M_{\sun}$, we see He-burning products at the surface, hence the zero values obtained for the N/C and N/O ratios. Only the non-rotating 40 $M_{\sun}$ model has H-burning products at the surface.

Comparing Fig.~\ref{FigHRDnorot} with Fig.~\ref{FigHRDrot}, we see that the portions of the tracks for which $\Delta \log(N/H)+12 > 10$ are more numerous and longer in the rotating models than in the non-rotating ones. This stage corresponds to a transition stage between the WN and the WC phases. When rotational mixing is accounted for, this transition is much smoother as already discussed by \citet{mm10}, and thus this stage is longer.

In Fig.~\ref{FigNHg}, the observed positions of the solar metallicity stars in the $\log(g)$ versus $\log(T_{\rm eff})$, and log(N/H)  versus $\log(g)$ planes are plotted (see the references for the observations in the caption of the figure). To see whether the models provide a good match to these observations, one must check that the track for an appropriate initial mass closely reproduces both the surface velocity and the surface enrichment at the appropriate evolutionary stage. Since for most of the stars only the $\upsilon \sin i$ is known (if it is known), we do not attempt any comparisons with the surface velocities. We merely assume that most of the stars should rotate with velocities near the observed averaged one for the B-type stars. In that case, the present rotating models should provide a reasonably good fit to the averaged surface enrichments. 

The left panel of Fig.~\ref{FigNHg} indicates that most of the sample stars have initial masses between 7 and 32 $M_{\sun}$. In the right panel of Fig.~\ref{FigNHg}, we see that globally, the theoretical models closely cover the ranges of observed abundances. We note that since the sample contains B-type stars, no observed points are plotted below $\log(g)\simeq 1$. To facilitate the comparison, we have coloured in red the observed stars with the highest N/H values. Interestingly, these stars are all evolved stars, and most of them cluster around the 32 $M_{\sun}$ track in the two panels of Fig.~\ref{FigNHg} giving some support to the present rotating models. 

For most of the observed points, the error bars in the N/H ratios are quite large (of the order of 0.20 dex). Only \citet{przybilla2010} reported smaller error bars of the order of 0.05 dex\footnote{Note that when the observed points of \citet{przybilla2010} are plotted in the N/C and N/O diagram they follow the trend expected based on the activity of the CNO cycle. Since this trend is not stellar-model dependent, but a feature resulting from nuclear physics, it provides a strong quality check of the surface abundances determinations.}. Second, many stars have smaller N/H values than the initial value  we adopted for our models. This might be because in some samples, stars of different initial metallicities, and thus different initial N/H ratios, have been observed. This again shows that precise and homogeneous samples are needed to provide constraints on the stellar models.

\begin{figure}
\centering
\includegraphics[width=.48\textwidth]{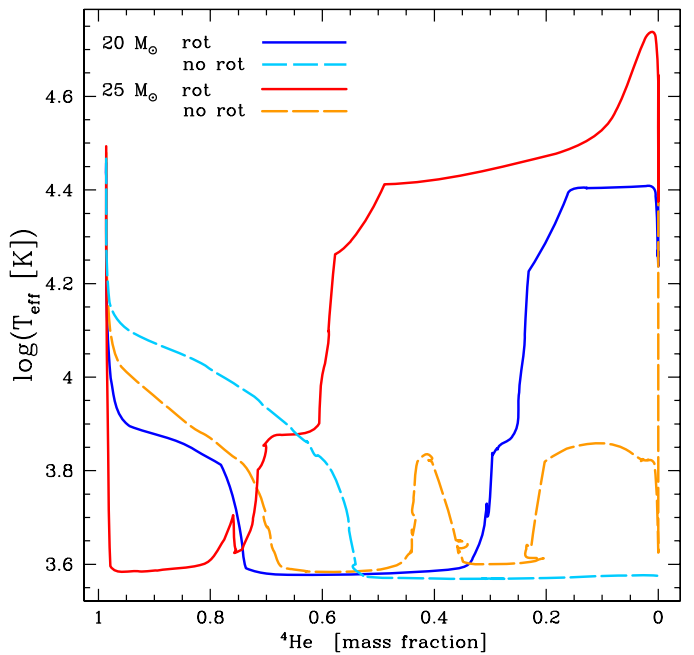}
 \caption{Evolutionary tracks of the 20 and 25 $M_{\sun}$ models in the diagram of the $\log(T_{\rm eff})$ as a function of the central helium mass fraction. In this diagram stars evolve from the left to the right during He-burning.}
      \label{BR}
\end{figure}
 
To test our models in a slightly different way, we have superimposed in Fig.~\ref{FigNHbis} (see the lower hatched areas) the mean values of the N/H enhancements for B-type dwarfs stars with masses below 20 $M_{\sun}$ as given in Table 2 of Maeder \& Meynet (2011, in press). These mean values are obtained from the observations by \citet{gieslamb92} and \citet{Kilian92} for a sample of 22 stars. The maximum observed values are shown in the upper hatched area. We see that the mean observed values are closely reproduced by our rotating models at the middle of the MS phase, which is quite encouraging. We can also see that the maximum values are reproduced well by the situation obtained in our models at the end of the MS phase. Some extreme cases can of course be found for very rapidly rotating models at an earlier stage, but the fraction of those stars is probably small (except maybe when regions rich in Be-type stars are observed).

Thus, we conclude from these comparisons that our models provide a good fit to the averaged observed surface enrichments.

\subsection{Some consequences for massive star evolution \label{SubSecMassiveEvol}}

We defer a detailed discussion of the Wolf-Rayet populations expected from these models to a forthcoming paper (Georgy et al., submitted). We discuss here only an interesting consequence of the present models for the evolution of stars passing through a red supergiant stage. As emphasized above, we implemented a new mass-loss prescription for those stars, which significantly enhances the mass-loss rate during this phase.  We note that this enhancement occurs only above a certain mass limit, which is between 15 and 20 $M_{\sun}$. This closely corresponds to the interval of masses containing the upper initial-mass limit of stars exploding as type II-P supernovae according to \citet{smartt09}. These authors indeed find that stars above a mass limit around 17-18 $M_{\sun}$ do not explode as type II-P supernovae, {\it i.e.} do not follow the characteristic lightcurve expected in the presence of a massive and extended H-rich envelope. Since red supergiants in the mass interval between 18 and about 30 $M_{\sun}$ are observed, one might wonder what kind of supernovae stars in this mass range would produce. Do they collapse to form a black hole, producing faint or even no supernova event? Or do they explode as type II-L, or type Ib supernova, as a result of the loss of the whole, or at least a great fraction, of their H-rich envelope? At the present stage, we would clearly favour this second hypothesis. There are indeed many observations \citep{smith04,vloon05} indicating that red supergiants (or at least part of them) experience stronger mass loss than presently accounted for in the models, which has led recently some authors \citep{YC2010} to explore the physical mechanisms responsible for this. In their work, they invoked the possibility that pulsation might trigger mass loss. Here, we propose another mechanism, that the luminosity reaches the Eddington luminosity in the outer layers. As indicated above, this implies that the enhancement will only occur above a given mass limit, which seems to be consistent with the finding of \citet{smartt09}.

To predict whether stellar models will explode as either a type II-P or a type II-L supernova remains however difficult.  We defer this discussion for a forthcoming paper, and only state that the present enhanced mass-loss rates favour the explosion as a type II-L or even a type Ib supernova.

Stronger mass loss during the red supergiant phase favours a bluewards evolution. This is a well known effect obtained many times in the past \citep[see for instance][]{salasnich99}. Thus, it changes the time spent in the blue and the red part of the HR diagram during the core He-burning phase. In Fig.~\ref{BR}, we see, as found by \citet{mm7}, that the first crossing of the Hertzsprung gap occurs more rapidly when rotation is accounted for. We also see that a large fraction of the end of the core He-burning phase (i.e., when Y $\lesssim$ 0.30-0.60) occurs in the blue part of the HR diagram in the rotating tracks. This is due to the combined effects of the enhanced mass-loss rates and rotational mixing, and has very important consequences for the blue to red supergiant ratio (B/R). In the models of \citet{schaller92}, the ratio of the blue to the red lifetime (red being defined by the time spent with an effective temperature below 3.65) for the 20 $M_{\sun}$ model is 0.2, while for the present non-rotating and rotating model, it is 0.7 and 1.5, respectively. The same numbers for the 25 $M_{\sun}$ models are respectively 0.10 \citep{schaller92}, 1.6 (present non-rotating), and 5.3 (rotating). Thus in that mass range, the B/R ratio is much larger with rotation. Whether it is sufficient to explain the B/R ratio observed in clusters remains to be studied and will be considered in a future paper. We note that at the metallicity of our models, the results found by \citet{eggenmm02} are between 1.2 and 4, in close agreement with our 20-25 $M_{\sun}$ rotating models.
 
Finally, this enhanced mass loss during the red supergiant phase lowers the minimum mass for single stars to become a WR star. \citet{schaller92} obtained an inferior mass limit of 32 $M_{\sun}$ \citep[see][]{meyn94}, while the present non-rotating and rotating models give 25 and 20 $M_{\sun}$ respectively. The consequences of this result for the WR populations at solar metallicity will be discussed in the next paper of this series.

\section{Discussion and conclusion \label{SecDiscu}}

We have studied for the first time, the impact of rotation on the evolution of single stars covering the mass domain from 0.8 to 120 $M_{\sun}$ in a homogeneous way. The present rotating models provide globally a good fit to the observed MS width, to the positions of the red giants and red supergiants in the HR diagram, and to the evolution of the surface velocities and the surface abundances. It is remarkable that such a good agreement can be obtained based on a unique well-chosen value of the initial $\upsilon/\upsilon_\text{crit}$ for most of the stellar mass range considered.

The rotating models provide a  more realistic view than non-rotating models in the sense that only models including rotational mixing are able to account for all the above features at the same time. Using a larger overshoot in non-rotating models, for instance, could reproduce the observed MS width but not the observed changes in the surface abundances and obviously not the surface velocities. All the data as well as the corresponding isochrones are made available through the web\footnote{\url{http://obswww.unige.ch/Recherche/evol/-Database-} or through the CDS database at  \url{http://vizier.u-strasbg.fr/viz-bin/VizieR-2}.}.

In addition to an extended grid of models with rotation, this work also illustrates the consequences of our revised mass-loss rate for the red supergiant phase. With this prescription, the stars between 20 and 32 $M_{\sun}$ lose very large amounts of mass, enabling some of them to evolve back to the blue part of the HR diagram, to change significantly the times spent in the blue and the red part of the HR diagram during the core He-burning phase, to lower the initial mass of stars that become WR stars and to change the lightcurve at the time of the supernova event.

In the future, we will provide grids for other initial metallicities. A very fine grid in mass for low-mass stars (without rotation) has been computed by Mowlavi et al. (submitted). This grid, computed with the same physics as here, allows very accurate interpolations for both age and mass determinations. 

\begin{acknowledgements}
The authors express their gratitude to Dr J. W. Ferguson, who has computed on request the molecular opacities for the peculiar mixture used in this paper. They also thank Claus Leitherer for his careful checking of the electronic files and preliminary work with the new models. R. H. acknowledges support from the World Premier International Research Center Initiative (WPI Initiative), MEXT, Japan. T.D. acknowledges support from the Swiss National Science Foundation (SNSF), from ESF-EuroGENESIS, and from the French Programme National de Physique Stellaire (PNPS) of CNRS/INSU.
\end{acknowledgements}
\bibliographystyle{aa}
\bibliography{BibTexRefs}
\end{document}